\documentclass[twocolumn, aps, prd, superscriptaddress,floatfix]{revtex4-1}

\usepackage{multirow}
\usepackage{amsmath,amssymb}
\usepackage{graphicx,bm}
\usepackage{slashed}
\usepackage{epstopdf}
\usepackage{ulem} 
\usepackage[usenames]{color}
\usepackage{float}
\usepackage{braket}
\usepackage{appendix}
\usepackage{hyperref}

\renewcommand\sout{\bgroup \color{red} \ULdepth=-.5ex \ULset}

\begin{document}

\title{Multibaryon configurations in a simple chromomagnetic model}

\author{Aaron Park}
\email{aaron.park@yonsei.ac.kr}\affiliation{Department of Physics and Institute of Physics and Applied Physics, Yonsei University, Seoul 03722, Korea}

\date{\today}
\begin{abstract}
In this work, we study the multibaryon configurations in a simple chromomagnetic model. We first construct the wave function of the multibaryon states using the multiquark configuration. We consider all possible quantum numbers assuming the spatial part of the wave function to be totally symmetric. Then, we calculate the color-spin factors for tetrabaryons, pentabaryons and hexabaryons in the flavor SU(3) breaking case.
\end{abstract}

\maketitle

\section{Introduction}
\label{Introduction}

In the realm of nuclear physics, the study of baryons has unveiled the intricate nature of the strong nuclear force that bind atomic nuclei together. 
In the past, much of the study on the interaction between baryons had focused on long range interactions driven by pion exchange. However, the interaction at a short distance could not be explained only with pion exchange, and for this, other explanations, such as vector meson exchange were needed. 

Among them, one of the most representative studies is a study examining short-range part of baryon-baryon interactions using the quark model \cite{Oka:1981ri,Oka:1981rj}. Moreover, we recently showed that the results of the quark model incorporating color-spin interaction are quite consistent with the results of the baryon-baryon interaction calculated in lattice QCD in the short range \cite{Park:2019bsz,Inoue:2016qxt}.

When dealing with baryon-baryon interactions in the quark model, the dibaryon configuration, which is a six-quark configuration is used. In particular, the task of calculating the color-spin factors of a dibaryon when the flavor SU(3) is broken has been studied frequently to examine the possibility of stable dibaryon as an exotic hadron. First, Silvestre-Brac and Leandri \cite{Silvestre-Brac:1992xsl} calculated the color-spin factors for all possible dibaryon states in the flavor SU(3) symmetry broken case. To describe the flavor symmetry broken effect, they introduced the following strange quark mass parameter.
\begin{align}
    \delta = 1 - \frac{m_u}{m_s},
\end{align}
where $m_u$ and $m_s$ are the constituent quark masses of $u,d$ and $s$, respectively. In this work, we represent the color-spin factors using this parameter.

Additionally, it has  been studied to examine the possibility of dibaryons containing heavy flavors using the constituent quark model \cite{Leandri:1993zg,Leandri:1995zm,Park:2016cmg,Park:2016mez,Richard:2020zxb,Huang:2020bmb,Liu:2022rzu}. If we only  consider $s$-wave state, the spatial part of the wave function should be totally symmetric. Then, more flavors increase the antisymmetry of the flavor state, and according to the Pauli principle, the remaining part of the wave function, which is the color-spin coupling state, becomes more symmetric making color-spin interaction attractive. Therefore, it is more likely that a compact exotic hadron containing heavy quarks exists than if it is composed of light quarks only. This can be seen as an another effect of color-spin interaction, which has a large effect in a short range.

Meanwhile, the behavior of nucleons in nuclei cannot be fully explained by two-body interactions alone. Three-body forces play a crucial role in refining our understanding of nuclear structure \cite{Fujita:1957zz,Epelbaum:2002vt,Hammer:2012id}.
Just as one studied two-body interaction using the dibaryon configuration, we have studied to analyze three-body interaction using the tribaryon configuration \cite{Park:2018ukx,Park:2019jff,Park:2020qpd}. Although the intrinsic three-body forces resulting from the color-spin interaction are canceled in the flavor SU(3) symmetric case \cite{Park:2019jff}, these can survive when the flavor symmetry is broken. Therefore, calculating the color-spin factor in the flavor symmetry broken case will be useful in studying the short range part of three-body interaction.

Additionally, we can go further considering four-baryon configuration. Tetrabaryon configuration may offer a unique opportunity to investigate the realm of four-body forces at short distances and their impact on nuclear dynamics. In this work, we calculate the color-spin factor of the tetrabaryon, pentabaryon, and hexabaryon in a simple chromomagnetic model.

This paper is organized as follows. In Sec.~\ref{Color-spin interaction section}, we introduce the color-spin interaction formula in the flavor SU(3) symmetric case. In Sec.~\ref{Tetrabryon configuration section}, we classify the possible tetrabaryon states and represent the color-spin factors for each cases when the flavor SU(3) symmetry is broken. Similalry, we represent the results for pentabaryons and hexabaryons in Sec.~\ref{Pentabaryon configuration section} and \ref{Hexabaryon configuration}. Finally, we summarize our work in Sec.~\ref{Summary section}.

\section{Color-spin interaction}
\label{Color-spin interaction section}

In this work, we construct the wave function of tetrabaryon, pentabaryon and hexabaryon assuming that the spatial part of the wave function to be totally symmetric. Then, we calculate the color-spin factor of multibaryon configurations using the following formula.
\begin{align}
V_{CS}&=-\sum_{i<j}^n \frac{1}{m_i m_j}\lambda^c_i \lambda^c_j \sigma_i \cdot \sigma_j \nonumber \\
&\equiv \frac{1}{m_u^2} H_{CS},
\label{color-spin}
\end{align}
where $\lambda^c_i$, $m_i$, $m_u$ are respectively the color SU(3) Gell-Mann matrices, the constituent quark mass of the $i$'th quark, and the constutient quark mass of $u,d$ quarks. 

In the flavor SU(3) symmetric case, color-spin factor $H_{CS}$ can be easily calculated by the following formula.

\begin{align}
  H_{CS}&=-\sum_{i<j}^n \lambda^c_i \lambda^c_j  \sigma_i \cdot \sigma_j \nonumber \\
  &= n(n-10) + \frac{4}{3}S(S+1)+4C_F+2C_C, \nonumber\\
  4C_F &= \frac{4}{3}(p_1^2+p_2^2+3p_1+3p_2+p_1 p_2),
\label{CSF-1}
\end{align}
where $C_F$ is the first kind of the Casimir operators of the flavor SU(3) and $p_i$ is the number of columns containing $i$ boxes in a column in the Young diagram.

\section{Tetrabaryon configuration}
\label{Tetrabryon configuration section}

In this section, we calculate the color-spin factors of the tetrabaryon which is four-baryon configuration. Since the color state of the tetrabryon is singlet, which is [4,4,4], we can determine the flavor-spin coupling state to be satisfied the Pauli exclusion principle as follows.

\begin{align}
    \begin{tabular}{|c|c|c|c|}
         \hline
         \quad \quad & \quad \quad & \quad \quad & \quad \quad  \\
         \hline
         \quad \quad & \quad \quad & \quad \quad & \quad \quad  \\
         \hline
         \quad \quad & \quad \quad & \quad \quad & \quad \quad  \\
         \hline
    \end{tabular}_C \otimes
    \begin{tabular}{|c|c|c|}
         \hline
         \quad \quad & \quad \quad & \quad \quad \\
         \hline
         \quad \quad & \quad \quad & \quad \quad \\
         \hline
         \quad \quad & \quad \quad & \quad \quad \\
         \hline
         \quad \quad & \quad \quad & \quad \quad \\
         \hline
    \end{tabular}_{FS}.
\end{align}
Now, we can decompose the flavor-spin coupling state into the possible flavor and spin states using Clebsh-Gordan(CG) series \cite{Stancu:1991rc}.\\

$[3,3,3,3]_{FS}=[6,6]_F \otimes [6,6]_S + [6,5,1]_F \otimes [7,5]_S + [6,4,2]_F \otimes [8,4]_S + [6,4,2]_F \otimes [6,6]_S + [6,3,3]_F \otimes [9,3]_S + [6,3,3]_F \otimes [7,5]_S + [5,5,2]_F \otimes [7,5]_S + [5,4,3]_F \otimes [8,4]_S + [5,4,3]_F \otimes [7,5]_S + [4,4,4]_F \otimes [6,6]_S $.\\

There are seven possible flavor states for tetrabaryon as follows. We indicate the possible spin states in the parentheses.

$\begin{tabular}{|c|c|c|c|}
  \cline{1-4}
  \quad \quad & \quad \quad & \quad \quad & \quad \quad \\
  \cline{1-4}
  \quad \quad & \quad \quad & \quad \quad & \quad \quad \\
  \cline{1-4}
  \quad \quad & \quad \quad & \quad \quad & \quad \quad \\
  \cline{1-4}
  \multicolumn{4}{c}{$\mathbf{1}(S=0)$}
\end{tabular}$,
$\begin{tabular}{|c|c|c|c|c|}
  \cline{1-5}
  \quad \quad & \quad \quad & \quad \quad & \quad \quad & \quad \quad \\
  \cline{1-5}
  \quad \quad & \quad \quad & \quad \quad & \quad \quad \\
  \cline{1-4}
  \quad \quad & \quad \quad & \quad \quad \\
  \cline{1-3}
  \multicolumn{5}{c}{$\mathbf{8}(S=1,2)$}
\end{tabular}$,
$\begin{tabular}{|c|c|c|c|c|}
  \cline{1-5}
  \quad \quad & \quad \quad & \quad \quad & \quad \quad & \quad \quad \\
  \cline{1-5}
  \quad \quad & \quad \quad & \quad \quad & \quad \quad & \quad \quad \\
  \cline{1-5}
  \quad \quad & \quad \quad \\
  \cline{1-2}
  \multicolumn{5}{c}{$\mathbf{\overline{10}}(S=1)$}
\end{tabular}$,
$\begin{tabular}{|c|c|c|c|c|c|}
  \cline{1-6}
  \quad \quad & \quad \quad & \quad \quad & \quad \quad & \quad \quad & \quad \quad \\
  \cline{1-6}
  \quad \quad & \quad \quad & \quad \quad  \\
  \cline{1-3}
  \quad \quad & \quad \quad & \quad \quad  \\
  \cline{1-3}
  \multicolumn{6}{c}{$\mathbf{10}(S=1,3)$}
\end{tabular}$,
$\begin{tabular}{|c|c|c|c|c|c|}
  \hline
  \quad \quad & \quad \quad & \quad \quad & \quad \quad & \quad \quad & \quad \quad \\
  \hline
  \quad \quad & \quad \quad & \quad \quad & \quad \quad \\
  \cline{1-4}
  \quad \quad & \quad \quad \\
  \cline{1-2}
  \multicolumn{6}{c}{$\mathbf{27}(S=0,2)$}
\end{tabular}$,\\
$\begin{tabular}{|c|c|c|c|c|c|}
  \hline
  \quad \quad & \quad \quad & \quad \quad & \quad \quad & \quad \quad & \quad \quad \\
  \hline
  \quad \quad & \quad \quad & \quad \quad & \quad \quad & \quad \quad \\
  \cline{1-5}
  \quad \quad \\
  \cline{1-1}
  \multicolumn{6}{c}{$\mathbf{\overline{35}}(S=1)$}
\end{tabular}$,
$\begin{tabular}{|c|c|c|c|c|c|}
  \hline
  \quad \quad & \quad \quad & \quad \quad & \quad \quad & \quad \quad & \quad \quad \\
  \hline
  \quad \quad & \quad \quad & \quad \quad & \quad \quad & \quad \quad & \quad \quad \\
  \hline
  \multicolumn{6}{c}{$\mathbf{\overline{28}}(S=0)$}
\end{tabular}$.

\begin{table}
\begin{tabular}{|c|c|c|c|c|c|c|c|c|}
  \hline
  \multirow{2}{*}{Flavor} & \multicolumn{4}{|c|}{$-\sum_{i<j} \lambda_i \lambda_j \sigma_i \cdot \sigma_j$} \\
  \cline{2-5}
   & $S=0$ & $S=1$ & $S=2$ & $S=3$ \\
  \hline
  $\mathbf{1}$ & $24$ & & &  \\
  \hline
  $\mathbf{8}$ &  & $\frac{116}{3}$ & 44 &  \\
  \hline
  $\mathbf{\overline{10}}$ &  & $\frac{152}{3}$ & & \\
  \hline
  $\mathbf{10}$ &  & $\frac{152}{3}$ & & 64 \\
  \hline
  $\mathbf{27}$ & 56 &  & 64 &  \\
  \hline
  $\mathbf{\overline{35}}$ &  & $\frac{224}{3}$ &  & \\
  \hline
  $\mathbf{\overline{28}}$ & 96 & & &  \\
  \hline
  \end{tabular}
\caption{The color-spin interaction factors of the tetrabaryon for each flavor and spin in the flavor SU(3) symmetric case. Blanks represent the Pauli blocking states.}
\label{tetrabaryon}
\end{table}

We represent the color-spin factors of the tetrabaryon in the flavor SU(3) symmetric case in Table \ref{tetrabaryon}. The results in the flavor SU(3) broken case are represented in the following subsections.

\subsection{$q^{12}$}

When there are no strange quarks, there is only one possible flavor which is $\mathbf{\overline{28}}$. Also, in this case, the possible isospin and spin are both zero.

\subsubsection{$I=0$}

$\begin{tabular}{|c|c|c|c|c|c|}
  \hline
  \quad \quad & \quad \quad & \quad \quad & \quad \quad & \quad \quad & \quad \quad \\
  \hline
  \quad \quad & \quad \quad & \quad \quad & \quad \quad & \quad \quad & \quad \quad \\
  \hline
  \multicolumn{6}{c}{$\mathbf{\overline{28}}(S=0)$}
\end{tabular}$.
\begin{align}
    H_{CS}=96.
\end{align}

\subsection{$q^{11}s: \{1,2,3,4,5,6,7,8,9,10,11\} 12$}

When there are strange quarks, the wave function must be constructed so that light quarks and strange quarks each satisfy the Pauli principle. So, in this case the wave function should satisfy \{1,2,3,4,5,6,7,8,9,10,11\} which means it is antisymmetric for exchange between eleven quarks. Also, among seven possible flavor states, only the flavor $\overline{\mathbf{35}}$ and $\overline{\mathbf{28}}$ states are possible for tetrabaryon with one strange quark. And in this case, only the isospin $\frac{1}{2}$ is allowed.

\subsubsection{$I=\frac{1}{2}$}

In this case, there are two possible flavor and spin states as follows.\\

$\begin{tabular}{|c|c|c|c|c|c|}
  \hline
  \quad \quad & \quad \quad & \quad \quad & \quad \quad & \quad \quad & \quad \quad \\
  \hline
  \quad \quad & \quad \quad & \quad \quad & \quad \quad & \quad \quad \\
  \cline{1-5}
  $\hspace{0.08cm} s\hspace{0.08cm}$ \\
  \cline{1-1}
  \multicolumn{6}{c}{$\mathbf{\overline{35}}(S=1)$}
\end{tabular}$,
$\begin{tabular}{|c|c|c|c|c|c|}
  \hline
  \quad \quad & \quad \quad & \quad \quad & \quad \quad & \quad \quad & \quad \quad \\
  \hline
  \quad \quad & \quad \quad & \quad \quad & \quad \quad & \quad \quad & $\hspace{0.08cm} s\hspace{0.08cm}$ \\
  \hline
  \multicolumn{6}{c}{$\mathbf{\overline{28}}(S=0)$}
\end{tabular}$.\\

For light quarks, color-spin coupling state is [2,2,2,2,2,1]. Since the color state of eleven quarks should be antitriplet, which is [4,4,3], we can find possible spin state using the CG series.

\begin{itemize}
    \item $S=1$ \\
    $[2,2,2,2,2,1]_{CS} = [4,4,3]_C \otimes [6,5]_S$. The spin of light quarks is $\frac{1}{2}$. Now, can determine the color-spin factor for light quarks.
    \begin{align}
        H^{uu}_{CS} =  80.
    \end{align}
    Then, we can determine the remaining color-spin factor because we know the total color-spin factor.
    \begin{align}
        H_{CS} &= \frac{224}{3} + \frac{16\delta}{3}.
    \end{align}
    However, it should be noted that this method is valid when there is only one possible state. If the multiplicity is two or more, then we need to calculate the color-spin factor starting from constructing the wave function.
    \item $S=0$ \\
    Similar to $S=1$ case, we can determine the color-spin factor.
    \begin{align}
        H_{CS} &= 96-16 \delta.
    \end{align}
\end{itemize}

\subsection{$q^{10}s^2: \{1,2,3,4,5,6,7,8,9,10 \} \{11,12\}$}

When there are two or more strange quarks, we should consider the symmetry between them too.
\subsubsection{$I=1$}

$F=(\begin{tabular}{|c|c|c|c|c|c|}
  \hline
  \quad \quad & \quad \quad & \quad \quad & \quad \quad & \quad \quad & \quad \quad \\
  \hline
  \quad \quad & \quad \quad & \quad \quad & \quad \quad \\
  \cline{1-4}
\end{tabular},
\begin{tabular}{|c|c|}
  \hline
  $\hspace{0.08cm} s\hspace{0.08cm}$ & $\hspace{0.08cm} s\hspace{0.08cm}$ \\
  \hline
\end{tabular})$, 
$CS=(\begin{tabular}{|c|c|}
  \hline
  \quad \quad & \quad \quad \\
  \hline
  \quad \quad & \quad \quad \\
  \hline
  \quad \quad & \quad \quad \\
  \hline
  \quad \quad & \quad \quad \\
  \hline
  \quad \quad \\
  \cline{1-1}
  \quad \quad \\
  \cline{1-1}
\end{tabular},
\begin{tabular}{|c|}
  \hline
  $\hspace{0.08cm} s\hspace{0.08cm}$ \\
  \hline
  $\hspace{0.08cm} s\hspace{0.08cm}$ \\
  \hline
\end{tabular})$\\

$[2,2,2,2,1,1]_{CS}=[4,4,2]_C \otimes [5,5]_S + [4,3,3]_C \otimes [6,4]_S$.\\

The possible flavor and spin states are as follows.

$\begin{tabular}{|c|c|c|c|c|c|}
  \hline
  \quad \quad & \quad \quad & \quad \quad & \quad \quad & \quad \quad & \quad \quad \\
  \hline
  \quad \quad & \quad \quad & \quad \quad & \quad \quad \\
  \cline{1-4}
  $\hspace{0.08cm} s\hspace{0.08cm}$ & $\hspace{0.08cm} s\hspace{0.08cm}$ \\
  \cline{1-2}
  \multicolumn{6}{c}{$\mathbf{27}(S=0,2)$}
\end{tabular}$,
$\begin{tabular}{|c|c|c|c|c|c|}
  \hline
  \quad \quad & \quad \quad & \quad \quad & \quad \quad & \quad \quad & \quad \quad \\
  \hline
  \quad \quad & \quad \quad & \quad \quad & \quad \quad & $\hspace{0.08cm} s\hspace{0.08cm}$ \\
  \cline{1-5}
  $\hspace{0.08cm} s\hspace{0.08cm}$ \\
  \cline{1-1}
  \multicolumn{6}{c}{$\mathbf{\overline{35}}(S=1)$}
\end{tabular}$,\\

$\begin{tabular}{|c|c|c|c|c|c|}
  \hline
  \quad \quad & \quad \quad & \quad \quad & \quad \quad & \quad \quad & \quad \quad \\
  \hline
  \quad \quad & \quad \quad & \quad \quad & \quad \quad & $\hspace{0.08cm} s\hspace{0.08cm}$ & $\hspace{0.08cm} s\hspace{0.08cm}$ \\
  \hline
  \multicolumn{6}{c}{$\mathbf{\overline{28}}(S=0)$}
\end{tabular}$.

\begin{itemize}
    \item $S=2$ \\
    $F=\mathbf{27}$
    \begin{align}
        H_{CS} &= 64 +\frac{8 \delta ^2}{3}.
    \end{align}
    \item $S=1$ \\
    $F=\mathbf{\overline{35}}$
    \begin{align}
        H_{CS} &= \frac{224}{3} -\frac{32 \delta }{3} + \frac{8 \delta ^2}{3}.
    \end{align}
    \item $S=0$ \\
    $F=\mathbf{\overline{28}}, \mathbf{27}$\\
    In this case, we need to construct the wave function of multiquark state. We represent the detailed Young tableaux showing the position of strange quarks in the appendix. 
    \begin{align}
        H_{CS} = \left(
                \begin{array}{cc}
                 96 -32 \delta + \frac{16 \delta ^2}{5}  & -\frac{4}{5} \sqrt{\frac{2}{3}}
                 \delta ^2 \\
                 -\frac{4}{5} \sqrt{\frac{2}{3}} \delta ^2 & 56 +8 \delta +\frac{52 \delta ^2}{15}
                  \\
\end{array}
\right).
    \end{align}
\end{itemize}

\subsubsection{$I=0$}

$F=(\begin{tabular}{|c|c|c|c|c|}
  \hline
  \quad \quad & \quad \quad & \quad \quad & \quad \quad & \quad \quad \\
  \hline
  \quad \quad & \quad \quad & \quad \quad & \quad \quad & \quad \quad \\
  \hline
\end{tabular},
\begin{tabular}{|c|c|}
  \hline
  $\hspace{0.08cm} s\hspace{0.08cm}$ & $\hspace{0.08cm} s\hspace{0.08cm}$ \\
  \hline
\end{tabular})$, 
$CS=(\begin{tabular}{|c|c|}
  \hline
  \quad \quad & \quad \quad \\
  \hline
  \quad \quad & \quad \quad \\
  \hline
  \quad \quad & \quad \quad \\
  \hline
  \quad \quad & \quad \quad \\
  \hline
  \quad \quad & \quad \quad \\
  \hline
\end{tabular},
\begin{tabular}{|c|}
  \hline
  $\hspace{0.08cm} s\hspace{0.08cm}$ \\
  \hline
  $\hspace{0.08cm} s\hspace{0.08cm}$ \\
  \hline
\end{tabular})$.\\

$[2,2,2,2,2]_{CS}=[4,4,2]_C \otimes [6,4]_S + [4,3,3]_C \otimes [5,5]_S$.\\

$\begin{tabular}{|c|c|c|c|c|}
  \cline{1-5}
  \quad \quad & \quad \quad & \quad \quad & \quad \quad & \quad \quad \\
  \cline{1-5}
  \quad \quad & \quad \quad & \quad \quad & \quad \quad & \quad \quad \\
  \cline{1-5}
  $\hspace{0.08cm} s\hspace{0.08cm}$ & $\hspace{0.08cm} s\hspace{0.08cm}$ \\
  \cline{1-2}
  \multicolumn{5}{c}{$\mathbf{\overline{10}}(S=1)$}
\end{tabular}$,
$\begin{tabular}{|c|c|c|c|c|c|}
  \hline
  \quad \quad & \quad \quad & \quad \quad & \quad \quad & \quad \quad & $\hspace{0.08cm} s\hspace{0.08cm}$ \\
  \hline
  \quad \quad & \quad \quad & \quad \quad & \quad \quad & \quad \quad \\
  \cline{1-5}
  $\hspace{0.08cm} s\hspace{0.08cm}$ \\
  \cline{1-1}
  \multicolumn{6}{c}{$\mathbf{\overline{35}}(S=1)$}
\end{tabular}$.

\begin{itemize}
    \item $S=1$ \\
    $F=\mathbf{\overline{35}},\mathbf{\overline{10}}$
    \begin{align}
        H_{CS} = \left(
\begin{array}{cc}
 \frac{224}{3}-\frac{160 \delta }{9}+\frac{32 \delta ^2}{9} & -\frac{16 \sqrt{2} \delta}{9}
   -\frac{4 \sqrt{2} \delta ^2}{9} \\
 -\frac{16 \sqrt{2} \delta}{9} -\frac{4 \sqrt{2} \delta ^2}{9}
   & \frac{152}{3}+\frac{40 \delta }{9}+\frac{28 \delta ^2}{9} \\
\end{array}
\right).
    \end{align}
\end{itemize}

\subsection{$q^9s^3: \{1,2,3,4,5,6,7,8,9 \} \{10,11,12\}$}

\subsubsection{$I=\frac{3}{2}$}

$F=(\begin{tabular}{|c|c|c|c|c|c|}
  \hline
  \quad \quad & \quad \quad & \quad \quad & \quad \quad & \quad \quad & \quad \quad \\
  \hline
  \quad \quad & \quad \quad & \quad \quad  \\
  \cline{1-3}
\end{tabular},
\begin{tabular}{|c|c|c|}
  \hline
  $\hspace{0.08cm} s\hspace{0.08cm}$ & $\hspace{0.08cm} s\hspace{0.08cm}$ & $\hspace{0.08cm} s\hspace{0.08cm}$ \\
  \hline
\end{tabular})$, \\
$CS=(\begin{tabular}{|c|c|}
  \hline
  \quad \quad & \quad \quad \\
  \hline
  \quad \quad & \quad \quad \\
  \hline
  \quad \quad & \quad \quad \\
  \hline
  \quad \quad \\
  \cline{1-1}
  \quad \quad \\
  \cline{1-1}
  \quad \quad \\
  \cline{1-1}
\end{tabular},
\begin{tabular}{|c|}
  \hline
  $\hspace{0.08cm} s\hspace{0.08cm}$ \\
  \hline
  $\hspace{0.08cm} s\hspace{0.08cm}$ \\
  \hline
  $\hspace{0.08cm} s\hspace{0.08cm}$ \\
  \hline
\end{tabular})$.\\

$[2,2,2,1,1,1]_{CS} = [4,3,2]_C \otimes [5,4]_S + [3,3,3]_C \otimes [6,3]_S$. \\

$\begin{tabular}{|c|c|c|c|c|c|}
  \cline{1-6}
  \quad \quad & \quad \quad & \quad \quad & \quad \quad & \quad \quad & \quad \quad \\
  \cline{1-6}
  \quad \quad & \quad \quad & \quad \quad  \\
  \cline{1-3}
  $\hspace{0.08cm} s\hspace{0.08cm}$ & $\hspace{0.08cm} s\hspace{0.08cm}$ & $\hspace{0.08cm} s\hspace{0.08cm}$  \\
  \cline{1-3}
  \multicolumn{6}{c}{$\mathbf{10}(S=1,3)$}
\end{tabular}$,
$\begin{tabular}{|c|c|c|c|c|c|}
  \hline
  \quad \quad & \quad \quad & \quad \quad & \quad \quad & \quad \quad & \quad \quad \\
  \hline
  \quad \quad & \quad \quad & \quad \quad & $\hspace{0.08cm} s\hspace{0.08cm}$ \\
  \cline{1-4}
  $\hspace{0.08cm} s\hspace{0.08cm}$ & $\hspace{0.08cm} s\hspace{0.08cm}$ \\
  \cline{1-2}
  \multicolumn{6}{c}{$\mathbf{27}(S=0,2)$}
\end{tabular}$,\\

$\begin{tabular}{|c|c|c|c|c|c|}
  \hline
  \quad \quad & \quad \quad & \quad \quad & \quad \quad & \quad \quad & \quad \quad \\
  \hline
  \quad \quad & \quad \quad & \quad \quad & $\hspace{0.08cm} s\hspace{0.08cm}$ & $\hspace{0.08cm} s\hspace{0.08cm}$ \\
  \cline{1-5}
  $\hspace{0.08cm} s\hspace{0.08cm}$ \\
  \cline{1-1}
  \multicolumn{6}{c}{$\mathbf{\overline{35}}(S=1)$}
\end{tabular}$,
$\begin{tabular}{|c|c|c|c|c|c|}
  \hline
  \quad \quad & \quad \quad & \quad \quad & \quad \quad & \quad \quad & \quad \quad \\
  \hline
  \quad \quad & \quad \quad & \quad \quad & $\hspace{0.08cm} s\hspace{0.08cm}$ & $\hspace{0.08cm} s\hspace{0.08cm}$ & $\hspace{0.08cm} s\hspace{0.08cm}$ \\
  \hline
  \multicolumn{6}{c}{$\mathbf{\overline{28}}(S=0)$}
\end{tabular}$.

\begin{itemize}
    \item $S=3$ \\
    $F=\mathbf{10}$
    \begin{align}
        H_{CS} &= 64-16 \delta +8 \delta ^2.
    \end{align}
    \item $S=2$ \\
    $F=\mathbf{27}$
    \begin{align}
        H_{CS} &= 64-16 \delta +8 \delta ^2.
    \end{align}
    \item $S=1$ \\
    $F=\mathbf{\overline{35}},\mathbf{10}$
    \begin{align}
        H_{CS} = \left(
\begin{array}{cc}
 \frac{224}{3}-\frac{80 \delta }{3}+\frac{80 \delta ^2}{9} & -\frac{4
   \sqrt{5} \delta ^2}{9}  \\
 -\frac{4 \sqrt{5} \delta ^2}{9}  & \frac{152}{3}-\frac{8 \delta
   }{3}+\frac{82 \delta ^2}{9} \\
\end{array}
\right).
    \end{align}
    \item $S=0$ \\
    $F=\mathbf{\overline{28}},\mathbf{27}$
    \begin{align}
        H_{CS} = \left(
\begin{array}{cc}
 96-48 \delta +\frac{48 \delta ^2}{5} & -\frac{4 \delta ^2}{5} 
   \\
 -\frac{4 \delta ^2}{5}  & 56-8 \delta +\frac{42 \delta ^2}{5}
   \\
\end{array}
\right).
    \end{align}
\end{itemize}

\subsubsection{$I=\frac{1}{2}$}

$F=(\begin{tabular}{|c|c|c|c|c|}
  \hline
  \quad \quad & \quad \quad & \quad \quad & \quad \quad & \quad \quad \\
  \hline
  \quad \quad & \quad \quad & \quad \quad & \quad \quad \\
  \cline{1-4}
\end{tabular},
\begin{tabular}{|c|c|c|}
  \hline
  $\hspace{0.08cm} s\hspace{0.08cm}$ & $\hspace{0.08cm} s\hspace{0.08cm}$ & $\hspace{0.08cm} s\hspace{0.08cm}$ \\
  \hline
\end{tabular})$, 
$CS=(\begin{tabular}{|c|c|}
  \hline
  \quad \quad & \quad \quad \\
  \hline
  \quad \quad & \quad \quad \\
  \hline
  \quad \quad & \quad \quad \\
  \hline
  \quad \quad & \quad \quad \\
  \hline
  \quad \quad \\
  \cline{1-1}
\end{tabular},
\begin{tabular}{|c|}
  \hline
  $\hspace{0.08cm} s\hspace{0.08cm}$ \\
  \hline
  $\hspace{0.08cm} s\hspace{0.08cm}$ \\
  \hline
  $\hspace{0.08cm} s\hspace{0.08cm}$ \\
  \hline
\end{tabular})$.\\

$[2,2,2,2,1]_{CS} = [4,4,1]_C \otimes [5,4]_S + [4,3,2]_C \otimes [6,3]_S + [4,3,2]_C \otimes [5,4]_S + [3,3,3]_C \otimes [5,4]_S$. \\

$\begin{tabular}{|c|c|c|c|c|}
  \cline{1-5}
  \quad \quad & \quad \quad & \quad \quad & \quad \quad & \quad \quad \\
  \cline{1-5}
  \quad \quad & \quad \quad & \quad \quad & \quad \quad \\
  \cline{1-4}
  $\hspace{0.08cm} s\hspace{0.08cm}$ & $\hspace{0.08cm} s\hspace{0.08cm}$ & $\hspace{0.08cm} s\hspace{0.08cm}$ \\
  \cline{1-3}
  \multicolumn{5}{c}{$\mathbf{8}(S=1,2)$}
\end{tabular}$,
$\begin{tabular}{|c|c|c|c|c|}
  \cline{1-5}
  \quad \quad & \quad \quad & \quad \quad & \quad \quad & \quad \quad \\
  \cline{1-5}
  \quad \quad & \quad \quad & \quad \quad & \quad \quad & $\hspace{0.08cm} s\hspace{0.08cm}$ \\
  \cline{1-5}
  $\hspace{0.08cm} s\hspace{0.08cm}$ & $\hspace{0.08cm} s\hspace{0.08cm}$ \\
  \cline{1-2}
  \multicolumn{5}{c}{$\mathbf{\overline{10}}(S=1)$}
\end{tabular}$,\\

$\begin{tabular}{|c|c|c|c|c|c|}
  \hline
  \quad \quad & \quad \quad & \quad \quad & \quad \quad & \quad \quad & $\hspace{0.08cm} s\hspace{0.08cm}$ \\
  \hline
  \quad \quad & \quad \quad & \quad \quad & \quad \quad \\
  \cline{1-4}
  $\hspace{0.08cm} s\hspace{0.08cm}$ & $\hspace{0.08cm} s\hspace{0.08cm}$ \\
  \cline{1-2}
  \multicolumn{6}{c}{$\mathbf{27}(S=0,2)$}
\end{tabular}$,
$\begin{tabular}{|c|c|c|c|c|c|}
  \hline
  \quad \quad & \quad \quad & \quad \quad & \quad \quad & \quad \quad & $\hspace{0.08cm} s\hspace{0.08cm}$ \\
  \hline
  \quad \quad & \quad \quad & \quad \quad & \quad \quad & $\hspace{0.08cm} s\hspace{0.08cm}$ \\
  \cline{1-5}
  $\hspace{0.08cm} s\hspace{0.08cm}$ \\
  \cline{1-1}
  \multicolumn{6}{c}{$\mathbf{\overline{35}}(S=1)$}
\end{tabular}$.

\begin{itemize}
    \item $S=2$  \\
    $F=\mathbf{27},\mathbf{8}$
    \begin{align}
        H_{CS} = \left(
\begin{array}{cc}
 64-\frac{128 \delta }{5}+\frac{48 \delta ^2}{5} & -\frac{16 \delta}{5} -\frac{4 \delta ^2}{5} \\
 -\frac{16 \delta}{5} -\frac{4 \delta ^2}{5} & 44-\frac{52 \delta
   }{5}+\frac{42 \delta ^2}{5} \\
\end{array}
\right).
    \end{align}
    \begin{widetext}
    \item $S=1$  \\
    $F=\mathbf{\overline{35}},\mathbf{\overline{10}},\mathbf{8}$
    \begin{align}
        H_{CS} = \left(
\begin{array}{ccc}
 \frac{224}{3}-\frac{112 \delta }{3}+\frac{88 \delta ^2}{9} & -\frac{8
   \delta}{3} -\frac{4 \delta ^2}{9} & -\frac{4 \delta ^2}{9} \\
 -\frac{8 \delta}{3} -\frac{4 \delta ^2}{9} & \frac{152}{3}-\frac{44
   \delta }{3}+\frac{82 \delta ^2}{9} & -\frac{8 \delta}{3} -\frac{8
   \delta ^2}{9} \\
 -\frac{4 \delta ^2}{9} & -\frac{8 \delta}{3} -\frac{8
   \delta ^2}{9} & \frac{116}{3}-4 \delta +\frac{82 \delta ^2}{9} \\
\end{array}
\right).
    \end{align}
    \end{widetext}
    \item $S=0$  \\
    $F=\mathbf{27}$
    \begin{align}
        H_{CS} &= 56-20 \delta +10 \delta ^2.
    \end{align}
\end{itemize}

\subsection{$q^8 s^4:\{1,2,3,4,5,6,7,8\} \{9,10,11,12\}$}

\subsubsection{$I=2$}

$F=(\begin{tabular}{|c|c|c|c|c|c|}
  \hline
  \quad \quad & \quad \quad & \quad \quad & \quad \quad & \quad \quad & \quad \quad \\
  \hline
  \quad \quad & \quad \quad \\
  \cline{1-2}
\end{tabular},
\begin{tabular}{|c|c|c|c|}
  \hline
  $\hspace{0.08cm} s\hspace{0.08cm}$ & $\hspace{0.08cm} s\hspace{0.08cm}$ & $\hspace{0.08cm} s\hspace{0.08cm}$ & $\hspace{0.08cm} s\hspace{0.08cm}$ \\
  \hline
\end{tabular})$, \\

$CS=(\begin{tabular}{|c|c|}
  \hline
  \quad \quad & \quad \quad \\
  \hline
  \quad \quad & \quad \quad \\
  \hline
  \quad \quad \\
  \cline{1-1}
  \quad \quad \\
  \cline{1-1}
  \quad \quad \\
  \cline{1-1}
  \quad \quad \\
  \cline{1-1}
\end{tabular},
\begin{tabular}{|c|}
  \hline
  $\hspace{0.08cm} s\hspace{0.08cm}$ \\
  \hline
  $\hspace{0.08cm} s\hspace{0.08cm}$ \\
  \hline
  $\hspace{0.08cm} s\hspace{0.08cm}$ \\
  \hline
  $\hspace{0.08cm} s\hspace{0.08cm}$ \\
  \hline
\end{tabular})$.\\

$[2,2,1,1,1,1]_{CS} = [4,2,2]_C \otimes [4,4]_S + [3,3,2]_C \otimes [5,3]_S $. \\

$\begin{tabular}{|c|c|c|c|c|c|}
  \hline
  \quad \quad & \quad \quad & \quad \quad & \quad \quad & \quad \quad & \quad \quad \\
  \hline
  \quad \quad & \quad \quad & $\hspace{0.08cm} s\hspace{0.08cm}$ & $\hspace{0.08cm} s\hspace{0.08cm}$ \\
  \cline{1-4}
  $\hspace{0.08cm} s\hspace{0.08cm}$ & $\hspace{0.08cm} s\hspace{0.08cm}$ \\
  \cline{1-2}
  \multicolumn{6}{c}{$\mathbf{27}(S=0,2)$}
\end{tabular}$,
$\begin{tabular}{|c|c|c|c|c|c|}
  \hline
  \quad \quad & \quad \quad & \quad \quad & \quad \quad & \quad \quad & \quad \quad \\
  \hline
  \quad \quad & \quad \quad & $\hspace{0.08cm} s\hspace{0.08cm}$ & $\hspace{0.08cm} s\hspace{0.08cm}$ & $\hspace{0.08cm} s\hspace{0.08cm}$ \\
  \cline{1-5}
  $\hspace{0.08cm} s\hspace{0.08cm}$ \\
  \cline{1-1}
  \multicolumn{6}{c}{$\mathbf{\overline{35}}(S=1)$}
\end{tabular}$,\\

$\begin{tabular}{|c|c|c|c|c|c|}
  \hline
  \quad \quad & \quad \quad & \quad \quad & \quad \quad & \quad \quad & \quad \quad \\
  \hline
  \quad \quad & \quad \quad & $\hspace{0.08cm} s\hspace{0.08cm}$ & $\hspace{0.08cm} s\hspace{0.08cm}$ & $\hspace{0.08cm} s\hspace{0.08cm}$ & $\hspace{0.08cm} s\hspace{0.08cm}$ \\
  \hline
  \multicolumn{6}{c}{$\mathbf{\overline{28}}(S=0)$}
\end{tabular}$.

\begin{itemize}
    \item $S=2$ \\
    $F=\mathbf{27}$
    \begin{align}
        H_{CS} &= 64-32 \delta +\frac{56 \delta ^2}{3}.
    \end{align}
    \item $S=1$ \\
    $F=\mathbf{\overline{35}}$
    \begin{align}
        H_{CS} &= \frac{224}{3}-\frac{128 \delta }{3}+\frac{56 \delta ^2}{3}.    
    \end{align}
    \item $S=0$ \\
    $F=\mathbf{\overline{28}},\mathbf{27}$
    \begin{align}
        H_{CS} = \left(
\begin{array}{cc}
 96-64 \delta +\frac{96 \delta ^2}{5} & -\frac{4}{5}  \sqrt{\frac{2}{3}}
   \delta ^2 \\
 -\frac{4}{5} \sqrt{\frac{2}{3}} \delta ^2 & 56-24 \delta +\frac{292
   \delta ^2}{15} \\
\end{array}
\right).
    \end{align}
\end{itemize}

\subsubsection{$I=1$}

$F=(\begin{tabular}{|c|c|c|c|c|}
  \hline
  \quad \quad & \quad \quad & \quad \quad & \quad \quad & \quad \quad \\
  \hline
  \quad \quad & \quad \quad & \quad \quad \\
  \cline{1-3}
\end{tabular},
\begin{tabular}{|c|c|c|c|}
  \hline
  $\hspace{0.08cm} s\hspace{0.08cm}$ & $\hspace{0.08cm} s\hspace{0.08cm}$ & $\hspace{0.08cm} s\hspace{0.08cm}$ & $\hspace{0.08cm} s\hspace{0.08cm}$ \\
  \hline
\end{tabular})$, \\

$CS=(\begin{tabular}{|c|c|}
  \hline
  \quad \quad & \quad \quad \\
  \hline
  \quad \quad & \quad \quad \\
  \hline
  \quad \quad & \quad \quad \\
  \hline
  \quad \quad \\
  \cline{1-1}
  \quad \quad \\
  \cline{1-1}
\end{tabular},
\begin{tabular}{|c|}
  \hline
  $\hspace{0.08cm} s\hspace{0.08cm}$ \\
  \hline
  $\hspace{0.08cm} s\hspace{0.08cm}$ \\
  \hline
  $\hspace{0.08cm} s\hspace{0.08cm}$ \\
  \hline
  $\hspace{0.08cm} s\hspace{0.08cm}$ \\
  \hline
\end{tabular}).$\\

$[2,2,2,1,1]_{CS} = [4,3,1]_C \otimes [5,3]_S + [4,3,1]_C \otimes [4,4]_S + [4,2,2]_C \otimes [5,3]_S + [3,3,2]_C \otimes [6,2]_S + [3,3,2]_C \otimes [5,3]_S + [3,3,2]_C \otimes [4,4]_S $. \\

$\begin{tabular}{|c|c|c|c|c|}
  \cline{1-5}
  \quad \quad & \quad \quad & \quad \quad & \quad \quad & \quad \quad \\
  \cline{1-5}
  \quad \quad & \quad \quad & \quad \quad & $\hspace{0.08cm} s\hspace{0.08cm}$ \\
  \cline{1-4}
  $\hspace{0.08cm} s\hspace{0.08cm}$ & $\hspace{0.08cm} s\hspace{0.08cm}$ & $\hspace{0.08cm} s\hspace{0.08cm}$ \\
  \cline{1-3}
  \multicolumn{5}{c}{$\mathbf{8}(S=1,2)$}
\end{tabular}$,
$\begin{tabular}{|c|c|c|c|c|}
  \cline{1-5}
  \quad \quad & \quad \quad & \quad \quad & \quad \quad & \quad \quad \\
  \cline{1-5}
  \quad \quad & \quad \quad & \quad \quad & $\hspace{0.08cm} s\hspace{0.08cm}$ & $\hspace{0.08cm} s\hspace{0.08cm}$ \\
  \cline{1-5}
  $\hspace{0.08cm} s\hspace{0.08cm}$ & $\hspace{0.08cm} s\hspace{0.08cm}$ \\
  \cline{1-2}
  \multicolumn{5}{c}{$\mathbf{\overline{10}}(S=1)$}
\end{tabular}$,\\

$\begin{tabular}{|c|c|c|c|c|c|}
  \cline{1-6}
  \quad \quad & \quad \quad & \quad \quad & \quad \quad & \quad \quad & $\hspace{0.08cm} s\hspace{0.08cm}$ \\
  \cline{1-6}
  \quad \quad & \quad \quad & \quad \quad  \\
  \cline{1-3}
  $\hspace{0.08cm} s\hspace{0.08cm}$ & $\hspace{0.08cm} s\hspace{0.08cm}$ & $\hspace{0.08cm} s\hspace{0.08cm}$  \\
  \cline{1-3}
  \multicolumn{6}{c}{$\mathbf{10}(S=1,3)$}
\end{tabular}$,
$\begin{tabular}{|c|c|c|c|c|c|}
  \hline
  \quad \quad & \quad \quad & \quad \quad & \quad \quad & \quad \quad & $\hspace{0.08cm} s\hspace{0.08cm}$ \\
  \hline
  \quad \quad & \quad \quad & \quad \quad & $\hspace{0.08cm} s\hspace{0.08cm}$ \\
  \cline{1-4}
  $\hspace{0.08cm} s\hspace{0.08cm}$ & $\hspace{0.08cm} s\hspace{0.08cm}$ \\
  \cline{1-2}
  \multicolumn{6}{c}{$\mathbf{27}(S=0,2)$}
\end{tabular}$,\\

$\begin{tabular}{|c|c|c|c|c|c|}
  \hline
  \quad \quad & \quad \quad & \quad \quad & \quad \quad & \quad \quad & $\hspace{0.08cm} s\hspace{0.08cm}$ \\
  \hline
  \quad \quad & \quad \quad & \quad \quad & $\hspace{0.08cm} s\hspace{0.08cm}$ & $\hspace{0.08cm} s\hspace{0.08cm}$ \\
  \cline{1-5}
  $\hspace{0.08cm} s\hspace{0.08cm}$ \\
  \cline{1-1}
  \multicolumn{6}{c}{$\mathbf{\overline{35}}(S=1)$}
\end{tabular}$.

\begin{itemize}
    \item $S=3$ \\
    $F=\mathbf{10}$
    \begin{align}
        H_{CS} &= 64-\frac{128 \delta }{3}+\frac{56 \delta ^2}{3}.
    \end{align}
    \item $S=2$ \\
    $F=\mathbf{27},\mathbf{8}$
    \begin{align}
        H_{CS} = \left(
\begin{array}{cc}
 64-\frac{224 \delta }{5}+\frac{56 \delta ^2}{3} & -\frac{16}{5} 
   \sqrt{\frac{2}{3}} \delta  \\
 -\frac{16}{5} \sqrt{\frac{2}{3}} \delta  & 44-\frac{388 \delta
   }{15}+\frac{56 \delta ^2}{3} \\
\end{array}
\right).
    \end{align}
    \begin{widetext}
    \item $S=1$ \\
    $F=\overline{\mathbf{10}},\mathbf{10},\mathbf{8},\overline{\mathbf{35}}$.\\
    In thise case, the calculation is a bit tricky. In most cases, we can calculate the color-spin factors using the CG coefficients of $S_n$, where $n$ is the number of nonstrange quarks. The reason why this is possible is because the values of the color-spin factor are different for each flavor. So, we can get the flavor eigenstates by diagonalization. However, in thise case flavor $\mathbf{10}$ and $\overline{\mathbf{10}}$ have the same color-spin factors, so we cannot distinguish which state is $\mathbf{10}$ and which state is $\overline{\mathbf{10}}$. And in fact, what we can obtain through diagonalization is the mixing of the two states. Therefore, in this case, we should construct the wave function using the CG coefficients of $S_{12}$ which clearly distinguish $\mathbf{10}$ and $\overline{\mathbf{10}}$.
    \begin{align}
        H_{CS} = \left(
\begin{array}{cccc}
 \frac{152}{3}-\frac{304 \delta }{9}+\frac{506 \delta ^2}{27} &
   -\frac{2 \sqrt{5}}{27}  \delta ^2 & \frac{8 \delta }{3}-\frac{4
   \delta ^2}{27} & \frac{16 \sqrt{2} \delta }{9}+\frac{4 \sqrt{2}
   \delta ^2}{27} \\
 -\frac{2 \sqrt{5}}{27}  \delta ^2 & \frac{152}{3}-\frac{304
   \delta }{9}+\frac{514 \delta ^2}{27} & \frac{8 \sqrt{5} \delta
   }{9}+\frac{4 \sqrt{5} \delta ^2}{27} & -\frac{4 \sqrt{10}}{27} 
   \delta ^2 \\
 \frac{8 \delta }{3}-\frac{4 \delta ^2}{27} & \frac{8 \sqrt{5} \delta
   }{9}+\frac{4 \sqrt{5} \delta ^2}{27} & \frac{116}{3}-\frac{188
   \delta }{9}+\frac{512 \delta ^2}{27} & -\frac{8 \sqrt{2}}{27} 
   \delta ^2 \\
 \frac{16 \sqrt{2} \delta }{9}+\frac{4 \sqrt{2} \delta ^2}{27} &
   -\frac{4 \sqrt{10}}{27}  \delta ^2 & -\frac{8 \sqrt{2}}{27} 
   \delta ^2 & \frac{224}{3}-\frac{512 \delta }{9}+\frac{520 \delta
   ^2}{27} \\
\end{array}
\right).
    \end{align}
    \end{widetext}
    \item $S=0$ \\
    $F=\mathbf{27}$
    \begin{align}
        H_{CS} &= 56-40 \delta +\frac{56 \delta ^2}{3}.
    \end{align}
\end{itemize}

\subsubsection{$I=0$}

$F=(\begin{tabular}{|c|c|c|c|}
  \hline
  \quad \quad & \quad \quad & \quad \quad & \quad \quad \\
  \hline
  \quad \quad & \quad \quad & \quad \quad & \quad \quad \\
  \hline
\end{tabular},
\begin{tabular}{|c|c|c|c|}
  \hline
  $\hspace{0.08cm} s\hspace{0.08cm}$ & $\hspace{0.08cm} s\hspace{0.08cm}$ & $\hspace{0.08cm} s\hspace{0.08cm}$ & $\hspace{0.08cm} s\hspace{0.08cm}$ \\
  \hline
\end{tabular})$, 
$CS=(\begin{tabular}{|c|c|}
  \hline
  \quad \quad & \quad \quad \\
  \hline
  \quad \quad & \quad \quad \\
  \hline
  \quad \quad & \quad \quad \\
  \hline
  \quad \quad & \quad \quad \\
  \hline
\end{tabular},
\begin{tabular}{|c|}
  \hline
  $\hspace{0.08cm} s\hspace{0.08cm}$ \\
  \hline
  $\hspace{0.08cm} s\hspace{0.08cm}$ \\
  \hline
  $\hspace{0.08cm} s\hspace{0.08cm}$ \\
  \hline
  $\hspace{0.08cm} s\hspace{0.08cm}$ \\
  \hline
\end{tabular})$.\\

$[2,2,2,2]_{CS} = [4,4]_C \otimes [4,4]_S + [4,3,1]_C \otimes [5,3]_S + [4,2,2]_C \otimes [6,2]_S + [4,2,2]_C \otimes [4,4]_S + [3,3,2]_C \otimes [5,3]_S$. \\

$\begin{tabular}{|c|c|c|c|}
  \cline{1-4}
  \quad \quad & \quad \quad & \quad \quad & \quad \quad \\
  \cline{1-4}
  \quad \quad & \quad \quad & \quad \quad & \quad \quad \\
  \cline{1-4}
  $\hspace{0.08cm} s\hspace{0.08cm}$ & $\hspace{0.08cm} s\hspace{0.08cm}$ & $\hspace{0.08cm} s\hspace{0.08cm}$ & $\hspace{0.08cm} s\hspace{0.08cm}$ \\
  \cline{1-4}
  \multicolumn{4}{c}{$\mathbf{1}(S=0)$}
\end{tabular}$,
$\begin{tabular}{|c|c|c|c|c|}
  \cline{1-5}
  \quad \quad & \quad \quad & \quad \quad & \quad \quad & $\hspace{0.08cm} s\hspace{0.08cm}$ \\
  \cline{1-5}
  \quad \quad & \quad \quad & \quad \quad & \quad \quad \\
  \cline{1-4}
  $\hspace{0.08cm} s\hspace{0.08cm}$ & $\hspace{0.08cm} s\hspace{0.08cm}$ & $\hspace{0.08cm} s\hspace{0.08cm}$ \\
  \cline{1-3}
  \multicolumn{5}{c}{$\mathbf{8}(S=1,2)$}
\end{tabular}$,
$\begin{tabular}{|c|c|c|c|c|c|}
  \hline
  \quad \quad & \quad \quad & \quad \quad & \quad \quad & $\hspace{0.08cm} s\hspace{0.08cm}$ & $\hspace{0.08cm} s\hspace{0.08cm}$ \\
  \hline
  \quad \quad & \quad \quad & \quad \quad & \quad \quad \\
  \cline{1-4}
  $\hspace{0.08cm} s\hspace{0.08cm}$ & $\hspace{0.08cm} s\hspace{0.08cm}$ \\
  \cline{1-2}
  \multicolumn{6}{c}{$\mathbf{27}(S=0,2)$}
\end{tabular}$.

\begin{itemize}
    \item $S=2$ \\
    $F=\mathbf{27},\mathbf{8}$
    \begin{align}
        H_{CS} = \left(
\begin{array}{cc}
 64-\frac{256 \delta }{5}+\frac{292 \delta ^2}{15} & -\frac{8
   \sqrt{6} \delta}{5} -\frac{4}{5} \sqrt{\frac{2}{3}} \delta ^2 \\
 -\frac{8 \sqrt{6} \delta}{5} -\frac{4}{5} \sqrt{\frac{2}{3}}
   \delta ^2 & 44-\frac{164 \delta }{5}+\frac{96 \delta ^2}{5} \\
\end{array}
\right).
    \end{align}
    \item $S=1$ \\
    $F=\mathbf{8}$
    \begin{align}
        H_{CS} &= \frac{116}{3}-\frac{92 \delta }{3}+\frac{56 \delta ^2}{3}.
    \end{align}
    \item $S=0$ \\
    $F=\mathbf{27},\mathbf{1}$
    \begin{align}
        H_{CS} = \left(
\begin{array}{cc}
 56-48 \delta +\frac{59 \delta ^2}{3} & -\frac{\delta ^2}{\sqrt{3}} \\
 -\frac{\delta ^2}{\sqrt{3}} & 24-16 \delta +19 \delta ^2 \\
\end{array}
\right).
    \end{align}
\end{itemize}

\subsection{$q^7 s^5:\{1,2,3,4,5,6,7\} \{8,9,10,11,12\}$}

\subsubsection{$I=\frac{5}{2}$}

$F=(\begin{tabular}{|c|c|c|c|c|c|}
  \hline
  \quad \quad & \quad \quad & \quad \quad & \quad \quad & \quad \quad & \quad \quad \\
  \hline
  \quad \quad  \\
  \cline{1-1}
\end{tabular},
\begin{tabular}{|c|c|c|c|c|}
  \hline
  $\hspace{0.08cm} s\hspace{0.08cm}$ & $\hspace{0.08cm} s\hspace{0.08cm}$ & $\hspace{0.08cm} s\hspace{0.08cm}$ & $\hspace{0.08cm} s\hspace{0.08cm}$ & $\hspace{0.08cm} s\hspace{0.08cm}$ \\
  \hline
\end{tabular})$, \\

$CS=(\begin{tabular}{|c|c|}
  \hline
  \quad \quad & \quad \quad \\
  \hline
  \quad \quad \\
  \cline{1-1}
  \quad \quad \\
  \cline{1-1}
  \quad \quad \\
  \cline{1-1}
  \quad \quad \\
  \cline{1-1}
  \quad \quad \\
  \cline{1-1}
\end{tabular},
\begin{tabular}{|c|}
  \hline
  $\hspace{0.08cm} s\hspace{0.08cm}$ \\
  \hline
  $\hspace{0.08cm} s\hspace{0.08cm}$ \\
  \hline
  $\hspace{0.08cm} s\hspace{0.08cm}$ \\
  \hline
  $\hspace{0.08cm} s\hspace{0.08cm}$ \\
  \hline
  $\hspace{0.08cm} s\hspace{0.08cm}$ \\
  \hline
\end{tabular})$.\\

$[2,1,1,1,1,1]_{CS} = [3,2,2]_C \otimes [4,3]_S $. \\

$\begin{tabular}{|c|c|c|c|c|c|}
  \hline
  \quad \quad & \quad \quad & \quad \quad & \quad \quad & \quad \quad & \quad \quad \\
  \hline
  \quad \quad & $\hspace{0.08cm} s\hspace{0.08cm}$ & $\hspace{0.08cm} s\hspace{0.08cm}$ & $\hspace{0.08cm} s\hspace{0.08cm}$ & $\hspace{0.08cm} s\hspace{0.08cm}$ \\
  \cline{1-5}
  $\hspace{0.08cm} s\hspace{0.08cm}$ \\
  \cline{1-1}
  \multicolumn{6}{c}{$\mathbf{\overline{35}}(S=1)$}
\end{tabular}$,
$\begin{tabular}{|c|c|c|c|c|c|}
  \hline
  \quad \quad & \quad \quad & \quad \quad & \quad \quad & \quad \quad & \quad \quad \\
  \hline
  \quad \quad & $\hspace{0.08cm} s\hspace{0.08cm}$ & $\hspace{0.08cm} s\hspace{0.08cm}$ & $\hspace{0.08cm} s\hspace{0.08cm}$ & $\hspace{0.08cm} s\hspace{0.08cm}$ & $\hspace{0.08cm} s\hspace{0.08cm}$ \\
  \hline
  \multicolumn{6}{c}{$\mathbf{\overline{28}}(S=0)$}
\end{tabular}$.

\begin{itemize}
    \item $S=1$ \\
    $F=\mathbf{\overline{35}}$
    \begin{align}
        H_{CS} &= \frac{224}{3}-\frac{176 \delta }{3}+32 \delta ^2.
    \end{align}
    \item $S=0$ \\
    $F=\mathbf{\overline{28}}$
    \begin{align}
        H_{CS} &= 96-80 \delta +32 \delta ^2.
    \end{align}
\end{itemize}

\subsubsection{$I=\frac{3}{2}$}

$F=(\begin{tabular}{|c|c|c|c|c|}
  \hline
  \quad \quad & \quad \quad & \quad \quad & \quad \quad & \quad \quad \\
  \hline
  \quad \quad & \quad \quad  \\
  \cline{1-2}
\end{tabular},
\begin{tabular}{|c|c|c|c|c|}
  \hline
  $\hspace{0.08cm} s\hspace{0.08cm}$ & $\hspace{0.08cm} s\hspace{0.08cm}$ & $\hspace{0.08cm} s\hspace{0.08cm}$ & $\hspace{0.08cm} s\hspace{0.08cm}$ & $\hspace{0.08cm} s\hspace{0.08cm}$ \\
  \hline
\end{tabular})$, \\

$CS=(\begin{tabular}{|c|c|}
  \hline
  \quad \quad & \quad \quad \\
  \hline
  \quad \quad & \quad \quad \\
  \hline
  \quad \quad \\
  \cline{1-1}
  \quad \quad \\
  \cline{1-1}
  \quad \quad \\
  \cline{1-1}
\end{tabular},
\begin{tabular}{|c|}
  \hline
  $\hspace{0.08cm} s\hspace{0.08cm}$ \\
  \hline
  $\hspace{0.08cm} s\hspace{0.08cm}$ \\
  \hline
  $\hspace{0.08cm} s\hspace{0.08cm}$ \\
  \hline
  $\hspace{0.08cm} s\hspace{0.08cm}$ \\
  \hline
  $\hspace{0.08cm} s\hspace{0.08cm}$ \\
  \hline
\end{tabular})$.\\

$[2,2,1,1,1]_{CS} = [4,2,1]_C \otimes [4,3]_S + [3,3,1]_C \otimes [5,2]_S + [3,3,1]_C \otimes [4,3]_S + [3,2,2]_C \otimes [5,2]_S + [3,2,2]_C \otimes [4,3]_S$. \\

$\begin{tabular}{|c|c|c|c|c|}
  \cline{1-5}
  \quad \quad & \quad \quad & \quad \quad & \quad \quad & \quad \quad \\
  \cline{1-5}
  \quad \quad & \quad \quad & $\hspace{0.08cm} s\hspace{0.08cm}$ & $\hspace{0.08cm} s\hspace{0.08cm}$ & $\hspace{0.08cm} s\hspace{0.08cm}$ \\
  \cline{1-5}
  $\hspace{0.08cm} s\hspace{0.08cm}$ & $\hspace{0.08cm} s\hspace{0.08cm}$ \\
  \cline{1-2}
  \multicolumn{5}{c}{$\mathbf{\overline{10}}(S=1)$}
\end{tabular}$,
$\begin{tabular}{|c|c|c|c|c|c|}
  \hline
  \quad \quad & \quad \quad & \quad \quad & \quad \quad & \quad \quad & $\hspace{0.08cm} s\hspace{0.08cm}$ \\
  \hline
  \quad \quad & \quad \quad & $\hspace{0.08cm} s\hspace{0.08cm}$ & $\hspace{0.08cm} s\hspace{0.08cm}$ \\
  \cline{1-4}
  $\hspace{0.08cm} s\hspace{0.08cm}$ & $\hspace{0.08cm} s\hspace{0.08cm}$ \\
  \cline{1-2}
  \multicolumn{6}{c}{$\mathbf{27}(S=0,2)$}
\end{tabular}$,\\

$\begin{tabular}{|c|c|c|c|c|c|}
  \hline
  \quad \quad & \quad \quad & \quad \quad & \quad \quad & \quad \quad & $\hspace{0.08cm} s\hspace{0.08cm}$ \\
  \hline
  \quad \quad & \quad \quad & $\hspace{0.08cm} s\hspace{0.08cm}$ & $\hspace{0.08cm} s\hspace{0.08cm}$ & $\hspace{0.08cm} s\hspace{0.08cm}$ \\
  \cline{1-5}
  $\hspace{0.08cm} s\hspace{0.08cm}$ \\
  \cline{1-1}
  \multicolumn{6}{c}{$\mathbf{\overline{35}}(S=1)$}
\end{tabular}$.

\begin{itemize}
    \item $S=2$ \\
    $F=\mathbf{27}$
    \begin{align}
        H_{CS} &= 64-64 \delta +32 \delta ^2.
    \end{align}
    \item $S=1$ \\
    $F=\mathbf{\overline{35}},\mathbf{\overline{10}}$
    \begin{align}
        H_{CS} = \left(
\begin{array}{cc}
 \frac{224}{3}-\frac{688 \delta }{9}+32 \delta ^2 & -\frac{8
   \sqrt{5} \delta}{9} \\
 -\frac{8 \sqrt{5} \delta}{9} & \frac{152}{3}-\frac{476
   \delta }{9}+32 \delta ^2 \\
\end{array}
\right).
    \end{align}
    \item $S=0$ \\
    $F=\mathbf{27}$
    \begin{align}
        H_{CS} &= 56-60 \delta +32 \delta ^2.
    \end{align}
\end{itemize}

\subsubsection{$I=\frac{1}{2}$}

$F=(\begin{tabular}{|c|c|c|c|}
  \hline
  \quad \quad & \quad \quad & \quad \quad & \quad \quad \\
  \hline
  \quad \quad & \quad \quad & \quad \quad \\
  \cline{1-3}
\end{tabular},
\begin{tabular}{|c|c|c|c|c|}
  \hline
  $\hspace{0.08cm} s\hspace{0.08cm}$ & $\hspace{0.08cm} s\hspace{0.08cm}$ & $\hspace{0.08cm} s\hspace{0.08cm}$ & $\hspace{0.08cm} s\hspace{0.08cm}$ & $\hspace{0.08cm} s\hspace{0.08cm}$ \\
  \hline
\end{tabular})$, \\

$CS=(\begin{tabular}{|c|c|}
  \hline
  \quad \quad & \quad \quad \\
  \hline
  \quad \quad & \quad \quad \\
  \hline
  \quad \quad & \quad \quad \\
  \hline
  \quad \quad \\
  \cline{1-1}
\end{tabular},
\begin{tabular}{|c|}
  \hline
  $\hspace{0.08cm} s\hspace{0.08cm}$ \\
  \hline
  $\hspace{0.08cm} s\hspace{0.08cm}$ \\
  \hline
  $\hspace{0.08cm} s\hspace{0.08cm}$ \\
  \hline
  $\hspace{0.08cm} s\hspace{0.08cm}$ \\
  \hline
  $\hspace{0.08cm} s\hspace{0.08cm}$ \\
  \hline
\end{tabular})$.\\

$[2,2,1,1,1]_{CS} = [4,3]_C \otimes [4,3]_S + [4,2,1]_C \otimes [5,2]_S + [4,2,1]_C \otimes [4,3]_S + [3,3,1]_C \otimes [5,2]_S + [3,3,1]_C \otimes [4,3]_S + [3,2,2]_C \otimes [6,1]_S + [3,2,2]_C \otimes [5,2]_S + [3,2,2]_C \otimes [4,3]_S$. \\

$\begin{tabular}{|c|c|c|c|c|}
  \cline{1-5}
  \quad \quad & \quad \quad & \quad \quad & \quad \quad & $\hspace{0.08cm} s\hspace{0.08cm}$ \\
  \cline{1-5}
  \quad \quad & \quad \quad & \quad \quad & $\hspace{0.08cm} s\hspace{0.08cm}$ \\
  \cline{1-4}
  $\hspace{0.08cm} s\hspace{0.08cm}$ & $\hspace{0.08cm} s\hspace{0.08cm}$ & $\hspace{0.08cm} s\hspace{0.08cm}$ \\
  \cline{1-3}
  \multicolumn{5}{c}{$\mathbf{8}(S=1,2)$}
\end{tabular}$,
$\begin{tabular}{|c|c|c|c|c|c|}
  \cline{1-6}
  \quad \quad & \quad \quad & \quad \quad & \quad \quad & $\hspace{0.08cm} s\hspace{0.08cm}$ & $\hspace{0.08cm} s\hspace{0.08cm}$ \\
  \cline{1-6}
  \quad \quad & \quad \quad & \quad \quad  \\
  \cline{1-3}
  $\hspace{0.08cm} s\hspace{0.08cm}$ & $\hspace{0.08cm} s\hspace{0.08cm}$ & $\hspace{0.08cm} s\hspace{0.08cm}$  \\
  \cline{1-3}
  \multicolumn{6}{c}{$\mathbf{10}(S=1,3)$}
\end{tabular}$,\\

$\begin{tabular}{|c|c|c|c|c|c|}
  \hline
  \quad \quad & \quad \quad & \quad \quad & \quad \quad & $\hspace{0.08cm} s\hspace{0.08cm}$ & $\hspace{0.08cm} s\hspace{0.08cm}$ \\
  \hline
  \quad \quad & \quad \quad & \quad \quad & $\hspace{0.08cm} s\hspace{0.08cm}$ \\
  \cline{1-4}
  $\hspace{0.08cm} s\hspace{0.08cm}$ & $\hspace{0.08cm} s\hspace{0.08cm}$ \\
  \cline{1-2}
  \multicolumn{6}{c}{$\mathbf{27}(S=0,2)$}
\end{tabular}$.

\begin{itemize}
    \item $S=3$ \\
    $F=\mathbf{10}$
    \begin{align}
        H_{CS} &= 64-\frac{208 \delta }{3}+32 \delta ^2.
    \end{align}
    \item $S=2$ \\
    $F=\mathbf{27},\mathbf{8}$
    \begin{align}
        H_{CS} = \left(
\begin{array}{cc}
 64-\frac{368 \delta }{5}+32 \delta ^2 & -\frac{16 \delta}{5}  \\
 -\frac{16 \delta}{5}  & 44-\frac{776 \delta }{15}+32 \delta ^2 \\
\end{array}
\right).
    \end{align}
    \item $S=1$ \\
    $F=\mathbf{10},\mathbf{8}$
    \begin{align}
        H_{CS} = \left(
\begin{array}{cc}
 \frac{152}{3}-\frac{584 \delta }{9}+32 \delta ^2 & -\frac{8
   \sqrt{5} \delta}{9}  \\
 -\frac{8
   \sqrt{5} \delta}{9} & \frac{116}{3}-\frac{472
   \delta }{9}+32 \delta ^2 \\
\end{array}
\right).
    \end{align}
    \item $S=0$ \\
    $F=\mathbf{27}$
    \begin{align}
        H_{CS} &= 56-72 \delta +32 \delta ^2.
    \end{align}
\end{itemize}

\subsection{$q^6 s^6:\{ 1,2,3,4,5,6 \} \{7,8,9,10,11,12 \}$}

In this case, the flavor and color-spin coupling state of strange quarks is as follows.\\

$\begin{tabular}{|c|c|c|c|c|c|}
  \hline
  $\hspace{0.08cm} s\hspace{0.08cm}$ & $\hspace{0.08cm} s\hspace{0.08cm}$ & $\hspace{0.08cm} s\hspace{0.08cm}$ & $\hspace{0.08cm} s\hspace{0.08cm}$ & $\hspace{0.08cm} s\hspace{0.08cm}$ & $\hspace{0.08cm} s\hspace{0.08cm}$ \\
  \hline
  \end{tabular}_F$ $\otimes$
$\begin{tabular}{|c|}
  \hline
  $\hspace{0.08cm} s\hspace{0.08cm}$   \\
  \hline
  $\hspace{0.08cm} s\hspace{0.08cm}$   \\
  \hline
  $\hspace{0.08cm} s\hspace{0.08cm}$   \\
  \hline
  $\hspace{0.08cm} s\hspace{0.08cm}$   \\
  \hline
  $\hspace{0.08cm} s\hspace{0.08cm}$   \\
  \hline
  $\hspace{0.08cm} s\hspace{0.08cm}$   \\
  \hline
\end{tabular}_{CS}$.\\

To satisfy the Pauli principle, decomposition of $[1,1,1,1,1,1]_{CS}$ for strange quarks should be as follows.\\

$\begin{tabular}{|c|c|}
  \hline
  $\hspace{0.08cm} s\hspace{0.08cm}$ & $\hspace{0.08cm} s\hspace{0.08cm}$ \\
  \hline
  $\hspace{0.08cm} s\hspace{0.08cm}$ & $\hspace{0.08cm} s\hspace{0.08cm}$ \\
  \hline
  $\hspace{0.08cm} s\hspace{0.08cm}$ & $\hspace{0.08cm} s\hspace{0.08cm}$ \\
  \hline
  \end{tabular}_C \otimes
  \begin{tabular}{|c|c|c|}
  \hline
  $\hspace{0.08cm} s\hspace{0.08cm}$ & $\hspace{0.08cm} s\hspace{0.08cm}$ & $\hspace{0.08cm} s\hspace{0.08cm}$ \\
  \hline
  $\hspace{0.08cm} s\hspace{0.08cm}$ & $\hspace{0.08cm} s\hspace{0.08cm}$ & $\hspace{0.08cm} s\hspace{0.08cm}$ \\
  \hline
  \end{tabular}_S$.\\

Therefore, the color-spin factor of strange quarks is determined regardless of the isospin of light quarks.
\begin{align}
    H^{ss}_{CS} = 48.
\end{align}
Additionally, we can find out that the color state of light quarks is also singlet because the color state of strange quarks is singlet.
In the same way, the spin of total quarks is determined from the spin of light quarks.

\subsubsection{$I=3$}

$\begin{tabular}{|c|c|c|c|c|c|}
  \hline
  \quad \quad & \quad \quad & \quad \quad & \quad \quad & \quad \quad & \quad \quad \\
  \hline
  $\hspace{0.08cm} s\hspace{0.08cm}$ & $\hspace{0.08cm} s\hspace{0.08cm}$ & $\hspace{0.08cm} s\hspace{0.08cm}$ & $\hspace{0.08cm} s\hspace{0.08cm}$ & $\hspace{0.08cm} s\hspace{0.08cm}$ & $\hspace{0.08cm} s\hspace{0.08cm}$ \\
  \hline
  \multicolumn{6}{c}{$\mathbf{\overline{28}}(S=0)$}
\end{tabular}$.\\

\begin{align}
    H_{CS} &= 96-96 \delta +48 \delta ^2.
\end{align}

\subsubsection{$I=2$}

$\begin{tabular}{|c|c|c|c|c|c|}
  \hline
  \quad \quad & \quad \quad & \quad \quad & \quad \quad & \quad \quad & $\hspace{0.08cm} s\hspace{0.08cm}$ \\
  \hline
  \quad \quad & $\hspace{0.08cm} s\hspace{0.08cm}$ & $\hspace{0.08cm} s\hspace{0.08cm}$ & $\hspace{0.08cm} s\hspace{0.08cm}$ & $\hspace{0.08cm} s\hspace{0.08cm}$ \\
  \cline{1-5}
  $\hspace{0.08cm} s\hspace{0.08cm}$ \\
  \cline{1-1}
  \multicolumn{6}{c}{$\mathbf{\overline{35}}(S=1)$}
\end{tabular}$.\\

\begin{align}
    H_{CS} &= \frac{224}{3}-96 \delta +48 \delta ^2.
\end{align}

\subsubsection{$I=1$}

$\begin{tabular}{|c|c|c|c|c|c|}
  \hline
  \quad \quad & \quad \quad & \quad \quad & \quad \quad & $\hspace{0.08cm} s\hspace{0.08cm}$ & $\hspace{0.08cm} s\hspace{0.08cm}$ \\
  \hline
  \quad \quad & \quad \quad & $\hspace{0.08cm} s\hspace{0.08cm}$ & $\hspace{0.08cm} s\hspace{0.08cm}$ \\
  \cline{1-4}
  $\hspace{0.08cm} s\hspace{0.08cm}$ & $\hspace{0.08cm} s\hspace{0.08cm}$ \\
  \cline{1-2}
  \multicolumn{6}{c}{$\mathbf{27}(S=0,2)$}
\end{tabular}$.

\begin{itemize}
    \item $S=2$\\ $F=\mathbf{27}$
    \begin{align}
        H_{CS} &= 64-96 \delta +48 \delta ^2.
    \end{align}
    \item $S=0$\\ $F=\mathbf{27}$
    \begin{align}
        H_{CS} &= 56-96 \delta +48 \delta ^2.
    \end{align}
\end{itemize}

\subsubsection{$I=0$}

$\begin{tabular}{|c|c|c|c|c|c|}
  \cline{1-6}
  \quad \quad & \quad \quad & \quad \quad & $\hspace{0.08cm} s\hspace{0.08cm}$ & $\hspace{0.08cm} s\hspace{0.08cm}$ & $\hspace{0.08cm} s\hspace{0.08cm}$ \\
  \cline{1-6}
  \quad \quad & \quad \quad & \quad \quad  \\
  \cline{1-3}
  $\hspace{0.08cm} s\hspace{0.08cm}$ & $\hspace{0.08cm} s\hspace{0.08cm}$ & $\hspace{0.08cm} s\hspace{0.08cm}$  \\
  \cline{1-3}
  \multicolumn{6}{c}{$\mathbf{10}(S=1,3)$}
\end{tabular}$.

\begin{itemize}
    \item $S=3$\\ $F=\mathbf{10}$
    \begin{align}
        H_{CS} &= 64-96 \delta +48 \delta ^2.
    \end{align}
    \item $S=1$\\ $F=\mathbf{10}$
    \begin{align}
        H_{CS} &= \frac{152}{3}-96 \delta +48 \delta ^2.
    \end{align}
\end{itemize}

\section{Pentabaryon configuration}
\label{Pentabaryon configuration section}

Now, let's consider the pentabaryon configuration which is composed of five baryons. Since the color state of the pentabaryon is singlet, which is [5,5,5], we can determine the flavor-spin coupling state satisfying the Pauli exclusion principle as follows.

\begin{align}
    \begin{tabular}{|c|c|c|c|c|}
         \hline
         \quad \quad & \quad \quad & \quad \quad & \quad \quad & \quad \quad \\
         \hline
         \quad \quad & \quad \quad & \quad \quad & \quad \quad & \quad \quad \\
         \hline
         \quad \quad & \quad \quad & \quad \quad & \quad \quad & \quad \quad \\
         \hline
    \end{tabular}_C \otimes
    \begin{tabular}{|c|c|c|}
         \hline
         \quad \quad & \quad \quad & \quad \quad \\
         \hline
         \quad \quad & \quad \quad & \quad \quad \\
         \hline
         \quad \quad & \quad \quad & \quad \quad \\
         \hline
         \quad \quad & \quad \quad & \quad \quad \\
         \hline
         \quad \quad & \quad \quad & \quad \quad \\
         \hline
    \end{tabular}_{FS}.
\end{align}
Now, we can decompose the flavor-spin coupling state into the possible flavor and spin states as follows.\\

$[3,3,3,3,3]_{FS}=[6,6,3]_F \otimes [9,6]_S + [6,5,4]_F \otimes [8,7]_S. $\\

There are two possible flavor states for the pentabaryon.\\

$\begin{tabular}{|c|c|c|c|c|c|}
  \hline
  \quad \quad & \quad \quad & \quad \quad & \quad \quad & \quad \quad & \quad \quad \\
  \hline
  \quad \quad & \quad \quad & \quad \quad & \quad \quad & \quad \quad & \quad \quad \\
  \hline
  \quad \quad & \quad \quad & \quad \quad \\
  \cline{1-3}
  \multicolumn{6}{c}{$\mathbf{\overline{10}}(S=\frac{3}{2})$}
\end{tabular}$,
$\begin{tabular}{|c|c|c|c|c|c|}
  \hline
  \quad \quad & \quad \quad & \quad \quad & \quad \quad & \quad \quad & \quad \quad \\
  \hline
  \quad \quad & \quad \quad & \quad \quad & \quad \quad & \quad \quad \\
  \cline{1-5}
  \quad \quad & \quad \quad & \quad \quad & \quad \quad \\
  \cline{1-4}
  \multicolumn{6}{c}{$\mathbf{8}(S=\frac{1}{2})$}
\end{tabular}$.\\

The color-spin factors for each state in the flavor symmetric case are as follows.
\begin{align}    H_{CS}^{\mathbf{\overline{10}}(S=\frac{3}{2})} &= 104 \\
    H_{CS}^{\mathbf{8}(S=\frac{1}{2})} &=  88
\end{align}

\subsection{$q^{12}s^3$}
According to possible flavor states of the pentabaryon, it should contain at least three strange quarks.

\subsubsection{$I=0$}

$\begin{tabular}{|c|c|c|c|c|c|}
  \hline
  \quad \quad & \quad \quad & \quad \quad & \quad \quad & \quad \quad & \quad \quad \\
  \hline
  \quad \quad & \quad \quad & \quad \quad & \quad \quad & \quad \quad & \quad \quad \\
  \hline
  $\hspace{0.08cm} s\hspace{0.08cm}$ & $\hspace{0.08cm} s\hspace{0.08cm}$ & $\hspace{0.08cm} s\hspace{0.08cm}$ \\
  \cline{1-3}
  \multicolumn{6}{c}{$\mathbf{\overline{10}}(S=\frac{3}{2})$}
\end{tabular}$.

\begin{itemize}
    \item $S=\frac{3}{2}$ \\ $F=\overline{\mathbf{10}}$
    \begin{align}
        H_{CS} = 104-16 \delta + 8\delta^2.
    \end{align}
\end{itemize}

\subsection{$q^{11}s^4$}

\subsubsection{$I=\frac{1}{2}$}

$\begin{tabular}{|c|c|c|c|c|c|}
  \hline
  \quad \quad & \quad \quad & \quad \quad & \quad \quad & \quad \quad & \quad \quad \\
  \hline
  \quad \quad & \quad \quad & \quad \quad & \quad \quad & \quad \quad & $\hspace{0.08cm} s\hspace{0.08cm}$ \\
  \hline
  $\hspace{0.08cm} s\hspace{0.08cm}$ & $\hspace{0.08cm} s\hspace{0.08cm}$ & $\hspace{0.08cm} s\hspace{0.08cm}$ \\
  \cline{1-3}
  \multicolumn{6}{c}{$\mathbf{\overline{10}}(S=\frac{3}{2})$}
\end{tabular}$,
$\begin{tabular}{|c|c|c|c|c|c|}
  \hline
  \quad \quad & \quad \quad & \quad \quad & \quad \quad & \quad \quad & \quad \quad \\
  \hline
  \quad \quad & \quad \quad & \quad \quad & \quad \quad & \quad \quad \\
  \cline{1-5}
  $\hspace{0.08cm} s\hspace{0.08cm}$ & $\hspace{0.08cm} s\hspace{0.08cm}$ & $\hspace{0.08cm} s\hspace{0.08cm}$ & $\hspace{0.08cm} s\hspace{0.08cm}$ \\
  \cline{1-4}
  \multicolumn{6}{c}{$\mathbf{8}(S=\frac{1}{2})$}
\end{tabular}$.

\begin{itemize}
    \item $S=\frac{3}{2}$ \\ $F=\overline{\mathbf{10}}$
    \begin{align}
        H_{CS} = 104-\frac{128}{3}\delta + \frac{56}{3}\delta^2.
    \end{align}
    \item $S=\frac{1}{2}$ \\ $F=\mathbf{8}$
    \begin{align}
        H_{CS} = 88-\frac{80}{3}\delta + \frac{56}{3}\delta^2.
    \end{align}
\end{itemize}

\subsection{$q^{10}s^5$}

\subsubsection{$I=1$}

$\begin{tabular}{|c|c|c|c|c|c|}
  \hline
  \quad \quad & \quad \quad & \quad \quad & \quad \quad & \quad \quad & \quad \quad \\
  \hline
  \quad \quad & \quad \quad & \quad \quad & \quad \quad & $\hspace{0.08cm} s\hspace{0.08cm}$ & $\hspace{0.08cm} s\hspace{0.08cm}$ \\
  \hline
  $\hspace{0.08cm} s\hspace{0.08cm}$ & $\hspace{0.08cm} s\hspace{0.08cm}$ & $\hspace{0.08cm} s\hspace{0.08cm}$ \\
  \cline{1-3}
  \multicolumn{6}{c}{$\mathbf{\overline{10}}(S=\frac{3}{2})$}
\end{tabular}$,
$\begin{tabular}{|c|c|c|c|c|c|}
  \hline
  \quad \quad & \quad \quad & \quad \quad & \quad \quad & \quad \quad & \quad \quad \\
  \hline
  \quad \quad & \quad \quad & \quad \quad & \quad \quad & $\hspace{0.08cm} s\hspace{0.08cm}$ \\
  \cline{1-5}
  $\hspace{0.08cm} s\hspace{0.08cm}$ & $\hspace{0.08cm} s\hspace{0.08cm}$ & $\hspace{0.08cm} s\hspace{0.08cm}$ & $\hspace{0.08cm} s\hspace{0.08cm}$ \\
  \cline{1-4}
  \multicolumn{6}{c}{$\mathbf{8}(S=\frac{1}{2})$}
\end{tabular}$.

\begin{itemize}
    \item $S=\frac{3}{2}$ \\ $F=\overline{\mathbf{10}}$
    \begin{align}
        H_{CS} = 104-\frac{208}{3}+32\delta^2.
    \end{align}
    \item $S=\frac{1}{2}$ \\ $F=\mathbf{8}$
    \begin{align}
        H_{CS} = 88-\frac{160}{3}\delta + 32\delta^2.
    \end{align}
\end{itemize}

\subsubsection{$I=0$}

$\begin{tabular}{|c|c|c|c|c|c|}
  \hline
  \quad \quad & \quad \quad & \quad \quad & \quad \quad & \quad \quad & $\hspace{0.08cm} s\hspace{0.08cm}$ \\
  \hline
  \quad \quad & \quad \quad & \quad \quad & \quad \quad & \quad \quad \\
  \cline{1-5}
  $\hspace{0.08cm} s\hspace{0.08cm}$ & $\hspace{0.08cm} s\hspace{0.08cm}$ & $\hspace{0.08cm} s\hspace{0.08cm}$ & $\hspace{0.08cm} s\hspace{0.08cm}$ \\
  \cline{1-4}
  \multicolumn{6}{c}{$\mathbf{8}(S=\frac{1}{2})$}
\end{tabular}$.

\begin{itemize}
    \item $S=\frac{1}{2}$ \\ $F=\mathbf{8}$
    \begin{align}
        H_{CS} = 88-64\delta + 32\delta^2.
    \end{align}
\end{itemize}

\subsection{$q^9s^6$}

\subsubsection{$I=\frac{3}{2}$}

$\begin{tabular}{|c|c|c|c|c|c|}
  \hline
  \quad \quad & \quad \quad & \quad \quad & \quad \quad & \quad \quad & \quad \quad \\
  \hline
  \quad \quad & \quad \quad & \quad \quad & $\hspace{0.08cm} s\hspace{0.08cm}$ & $\hspace{0.08cm} s\hspace{0.08cm}$ & $\hspace{0.08cm} s\hspace{0.08cm}$ \\
  \hline
  $\hspace{0.08cm} s\hspace{0.08cm}$ & $\hspace{0.08cm} s\hspace{0.08cm}$ & $\hspace{0.08cm} s\hspace{0.08cm}$ \\
  \cline{1-3}
  \multicolumn{6}{c}{$\mathbf{\overline{10}}(S=\frac{3}{2})$}
\end{tabular}$.

\begin{itemize}
    \item $S=\frac{3}{2}$ \\ $F=\overline{\mathbf{10}}$
    \begin{align}
        H_{CS} = 104-96\delta + 48\delta^2.
    \end{align}
\end{itemize}

\subsubsection{$I=\frac{1}{2}$}

$\begin{tabular}{|c|c|c|c|c|c|}
  \hline
  \quad \quad & \quad \quad & \quad \quad & \quad \quad & \quad \quad & $\hspace{0.08cm} s\hspace{0.08cm}$ \\
  \hline
  \quad \quad & \quad \quad & \quad \quad & \quad \quad & $\hspace{0.08cm} s\hspace{0.08cm}$ \\
  \cline{1-5}
  $\hspace{0.08cm} s\hspace{0.08cm}$ & $\hspace{0.08cm} s\hspace{0.08cm}$ & $\hspace{0.08cm} s\hspace{0.08cm}$ & $\hspace{0.08cm} s\hspace{0.08cm}$ \\
  \cline{1-4}
  \multicolumn{6}{c}{$\mathbf{8}(S=\frac{1}{2})$}
\end{tabular}$.

\begin{itemize}
    \item $S=\frac{1}{2}$ \\ $F=\mathbf{8}$
    \begin{align}
        H_{CS} = 88-96\delta + 48\delta^2.
    \end{align}
\end{itemize}

\section{Hexabaryon configuration}
\label{Hexabaryon configuration}

We can also consider the  hexabaryon configuration which is composed of six baryons. Since the color state of the hexabaryon is singlet, which is [6,6,6], we can determine the flavor-spin coupling state satisfying the Pauli exclusion principle as follows.

\begin{align}
    \begin{tabular}{|c|c|c|c|c|c|}
         \hline
         \quad \quad & \quad \quad & \quad \quad & \quad \quad & \quad \quad & \quad \quad \\
         \hline
         \quad \quad & \quad \quad & \quad \quad & \quad \quad & \quad \quad & \quad \quad \\
         \hline
         \quad \quad & \quad \quad & \quad \quad & \quad \quad & \quad \quad & \quad \quad \\
         \hline
    \end{tabular}_C \otimes
    \begin{tabular}{|c|c|c|}
         \hline
         \quad \quad & \quad \quad & \quad \quad \\
         \hline
         \quad \quad & \quad \quad & \quad \quad \\
         \hline
         \quad \quad & \quad \quad & \quad \quad \\
         \hline
         \quad \quad & \quad \quad & \quad \quad \\
         \hline
         \quad \quad & \quad \quad & \quad \quad \\
         \hline
         \quad \quad & \quad \quad & \quad \quad \\
         \hline
    \end{tabular}_{FS}.
\end{align}
Now, we can decompose the flavor-spin coupling state into the possible flavor and spin states as follows.\\

$[3,3,3,3,3,3]_{FS}=[6,6,6]_F \otimes [9,9]_S. $\\

There is one possible flavor state for the hexabaryon.

$\begin{tabular}{|c|c|c|c|c|c|}
  \hline
  \quad \quad & \quad \quad & \quad \quad & \quad \quad & \quad \quad & \quad \quad \\
  \hline
  \quad \quad & \quad \quad & \quad \quad & \quad \quad & \quad \quad & \quad \quad \\
  \hline
  \quad \quad & \quad \quad & \quad \quad & \quad \quad & \quad \quad & \quad \quad \\
  \hline
  \multicolumn{6}{c}{$\mathbf{1}(S=0)$}
\end{tabular}$.\\

The color-spin factor for this state in the flavor symmetric case is as follows.
\begin{align}
    H_{CS} = 144.
\end{align}

\subsection{$q^{12}s^6$}

For the flavor singlet of the hexabaryon, it should contain six strange quarks.
\subsubsection{$I=0$}

$\begin{tabular}{|c|c|c|c|c|c|}
  \hline
  \quad \quad & \quad \quad & \quad \quad & \quad \quad & \quad \quad & \quad \quad \\
  \hline
  \quad \quad & \quad \quad & \quad \quad & \quad \quad & \quad \quad & \quad \quad \\
  \hline
  $\hspace{0.08cm} s\hspace{0.08cm}$ & $\hspace{0.08cm} s\hspace{0.08cm}$ & $\hspace{0.08cm} s\hspace{0.08cm}$ & $\hspace{0.08cm} s\hspace{0.08cm}$ & $\hspace{0.08cm} s\hspace{0.08cm}$ & $\hspace{0.08cm} s\hspace{0.08cm}$ \\
  \hline
  \multicolumn{6}{c}{$\mathbf{1}(S=0)$}
\end{tabular}$.

\begin{align}
    H_{CS} = 144 - 96\delta + 48\delta^2.
\end{align}

\section{Summary}
\label{Summary section}

In this work, we calculated the matrix elements of color-spin factors of the multibaryon configurations, which are tetrabaryons, pentabaryons and hexabaryons, in flavor SU(3) symmetry broken case. First, we constructed the wave function of the multibaryon states assuming the spatial part to be totally symmetric. Since the color state of the multibaryon should be singlet, we could determine the remaining part of the wave function which is flavor-spin coupling state satisfying the Pauli exclusion principle. Then, using the CG series we decomposed the flavor-spin coupling state into flavor and spin states, respectively. And then, we  calculated the color-spin factors using the CG coefficients of $S_n$, where $n$ is the number of nonstrange quarks, except the $q^8 s^4(I=1,S=1)$ case.

As our previous work showed \cite{Park:2019bsz}, color-spin interaction could be critical ingredient when we study the short range part of the baryon-baryon interaction. In dense nuclear matter, three-body and even four-body interations can be more important. Therefore, the color-spin factors calculated in this work may be useful when studying extremely high density nuclear matter.

However, it should be noted that in this work, we assumed that the spatial part of the wave function to be totally symmetric. In this case, the six-baryon is the maximum multibaryon state we can construct to be satisfied the Pauli exclusion principle. Moreover, since it is important to consider the angular momentum in three-body or larger systems, follow-up research considering the nonsymmetric orbital state is  necessary.

\begin{widetext}

\appendix

\section{$q^{10} s^2(I=1,S=0)$}

In the following sections, we represent how strange quarks occupy their places in color and spin states for the cases when there are two or more multiplicities. Note that the color-spin coupling state of strange quarks should be antisymmetric.\\

1. $\begin{tabular}{|c|c|c|c|}
  \hline
  \quad \quad & \quad \quad & \quad \quad & \quad \quad \\
  \hline
  \quad \quad & \quad \quad & \quad \quad & \quad \quad \\
  \hline
  \quad \quad & \quad \quad & $\hspace{0.08cm} s\hspace{0.08cm}$ & $\hspace{0.08cm} s\hspace{0.08cm}$ \\
  \hline
\end{tabular}_C \otimes$
$\begin{tabular}{|c|c|c|c|c|c|}
  \hline
  \quad \quad & \quad \quad & \quad \quad & \quad \quad & \quad \quad & $\hspace{0.08cm} s\hspace{0.08cm}$ \\
  \hline
  \quad \quad & \quad \quad & \quad \quad & \quad \quad & \quad \quad & $\hspace{0.08cm} s\hspace{0.08cm}$ \\
  \hline
\end{tabular}_S \rightarrow
\left( \begin{tabular}{|c|c|}
  \hline
  \quad \quad & \quad \quad \\
  \hline
  \quad \quad & \quad \quad \\
  \hline
  \quad \quad & \quad \quad \\
  \hline
  \quad \quad & \quad \quad \\
  \hline
  \quad \quad \\
  \cline{1-1}
  \quad \quad \\
  \cline{1-1}
\end{tabular},
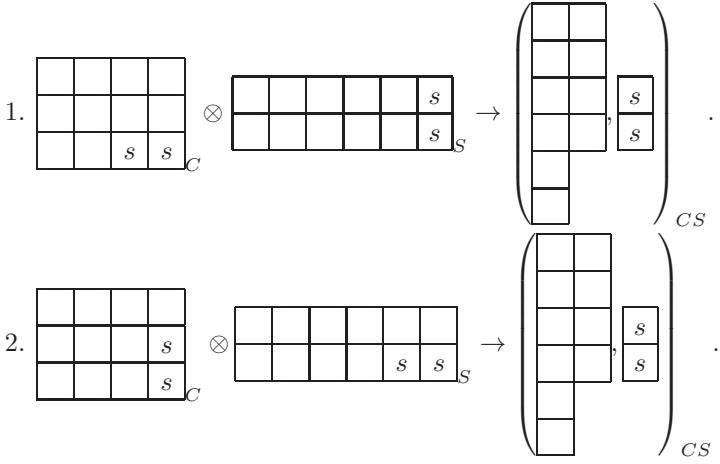
\begin{tabular}{|c|}
  \hline
  $\hspace{0.08cm} s\hspace{0.08cm}$ \\
  \hline
  $\hspace{0.08cm} s\hspace{0.08cm}$ \\
  \hline
\end{tabular}
 \right)_{CS}$. \\

 2. $\begin{tabular}{|c|c|c|c|}
  \hline
  \quad \quad & \quad \quad & \quad \quad & \quad \quad \\
  \hline
  \quad \quad & \quad \quad & \quad \quad & $\hspace{0.08cm} s\hspace{0.08cm}$ \\
  \hline
  \quad \quad & \quad \quad & \quad \quad & $\hspace{0.08cm} s\hspace{0.08cm}$ \\
  \hline
\end{tabular}_C \otimes
\begin{tabular}{|c|c|c|c|c|c|}
  \hline
  \quad \quad & \quad \quad & \quad \quad & \quad \quad & \quad \quad & \quad \quad \\
  \hline
  \quad \quad & \quad \quad & \quad \quad & \quad \quad & $\hspace{0.08cm} s\hspace{0.08cm}$ & $\hspace{0.08cm} s\hspace{0.08cm}$ \\
  \hline
\end{tabular}_S \rightarrow
\left(\begin{tabular}{|c|c|}
  \hline
  \quad \quad & \quad \quad \\
  \hline
  \quad \quad & \quad \quad \\
  \hline
  \quad \quad & \quad \quad \\
  \hline
  \quad \quad & \quad \quad \\
  \hline
  \quad \quad \\
  \cline{1-1}
  \quad \quad \\
  \cline{1-1}
\end{tabular},
\begin{tabular}{|c|}
  \hline
  $\hspace{0.08cm} s\hspace{0.08cm}$ \\
  \hline
  $\hspace{0.08cm} s\hspace{0.08cm}$ \\
  \hline
\end{tabular}
 \right)_{CS}$.

 \section{$q^{10} s^2(I=0,S=1)$}

1. $\begin{tabular}{|c|c|c|c|}
  \hline
  \quad \quad & \quad \quad & \quad \quad & \quad \quad \\
  \hline
  \quad \quad & \quad \quad & \quad \quad & \quad \quad \\
  \hline
  \quad \quad & \quad \quad & $\hspace{0.08cm} s\hspace{0.08cm}$ & $\hspace{0.08cm} s\hspace{0.08cm}$ \\
  \hline
\end{tabular}_C \otimes$
$\begin{tabular}{|c|c|c|c|c|c|c|}
  \hline
  \quad \quad & \quad \quad & \quad \quad & \quad \quad & \quad \quad & \quad \quad  & $\hspace{0.08cm} s\hspace{0.08cm}$ \\
  \hline
  \quad \quad & \quad \quad & \quad \quad & \quad \quad   & $\hspace{0.08cm} s\hspace{0.08cm}$ \\
  \cline{1-5}
\end{tabular}_S \rightarrow
\left(\begin{tabular}{|c|c|}
  \hline
  \quad \quad & \quad \quad \\
  \hline
  \quad \quad & \quad \quad \\
  \hline
  \quad \quad & \quad \quad \\
  \hline
  \quad \quad & \quad \quad \\
  \hline
  \quad \quad & \quad \quad \\
  \hline
\end{tabular},
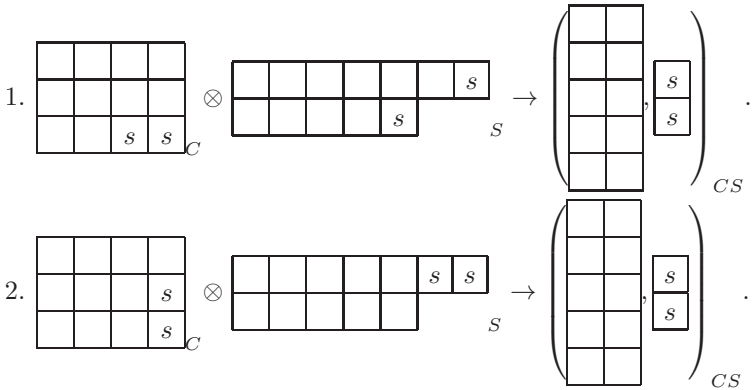
\begin{tabular}{|c|}
  \hline
  $\hspace{0.08cm} s\hspace{0.08cm}$ \\
  \hline
  $\hspace{0.08cm} s\hspace{0.08cm}$ \\
  \hline
\end{tabular}
 \right)_{CS}$. \\

 2. $\begin{tabular}{|c|c|c|c|}
  \hline
  \quad \quad & \quad \quad & \quad \quad & \quad \quad \\
  \hline
  \quad \quad & \quad \quad & \quad \quad & $\hspace{0.08cm} s\hspace{0.08cm}$ \\
  \hline
  \quad \quad & \quad \quad & \quad \quad & $\hspace{0.08cm} s\hspace{0.08cm}$ \\
  \hline
\end{tabular}_C \otimes$
$\begin{tabular}{|c|c|c|c|c|c|c|}
  \hline
  \quad \quad & \quad \quad & \quad \quad & \quad \quad & \quad \quad & $\hspace{0.08cm} s\hspace{0.08cm}$ & $\hspace{0.08cm} s\hspace{0.08cm}$ \\
  \hline
  \quad \quad & \quad \quad & \quad \quad & \quad \quad & \quad \quad \\
  \cline{1-5}
\end{tabular}_S \rightarrow
\left(\begin{tabular}{|c|c|}
  \hline
  \quad \quad & \quad \quad \\
  \hline
  \quad \quad & \quad \quad \\
  \hline
  \quad \quad & \quad \quad \\
  \hline
  \quad \quad & \quad \quad \\
  \hline
  \quad \quad & \quad \quad \\
  \hline
\end{tabular},
\begin{tabular}{|c|}
  \hline
  $\hspace{0.08cm} s\hspace{0.08cm}$ \\
  \hline
  $\hspace{0.08cm} s\hspace{0.08cm}$ \\
  \hline
\end{tabular}
 \right)_{CS}$.

\section{$q^{9} s^3(I=\frac{3}{2},S=1)$}

1. $\begin{tabular}{|c|c|c|c|}
  \hline
  \quad \quad & \quad \quad & \quad \quad & \quad \quad \\
  \hline
  \quad \quad & \quad \quad & \quad \quad & $\hspace{0.08cm} s\hspace{0.08cm}$ \\
  \hline
  \quad \quad & \quad \quad & $\hspace{0.08cm} s\hspace{0.08cm}$ & $\hspace{0.08cm} s\hspace{0.08cm}$ \\
  \hline
\end{tabular}_C \otimes$
$\begin{tabular}{|c|c|c|c|c|c|c|}
  \hline
  \quad \quad & \quad \quad & \quad \quad & \quad \quad & \quad \quad & $\hspace{0.08cm} s\hspace{0.08cm}$  & $\hspace{0.08cm} s\hspace{0.08cm}$ \\
  \hline
  \quad \quad & \quad \quad & \quad \quad & \quad \quad   & $\hspace{0.08cm} s\hspace{0.08cm}$ \\
  \cline{1-5}
\end{tabular}_S \rightarrow$
$\left(\begin{tabular}{|c|c|}
  \hline
  \quad \quad & \quad \quad \\
  \hline
  \quad \quad & \quad \quad \\
  \hline
  \quad \quad & \quad \quad \\
  \hline
  \quad \quad \\
  \cline{1-1}
  \quad \quad \\
  \cline{1-1}
  \quad \quad \\
  \cline{1-1}
\end{tabular},
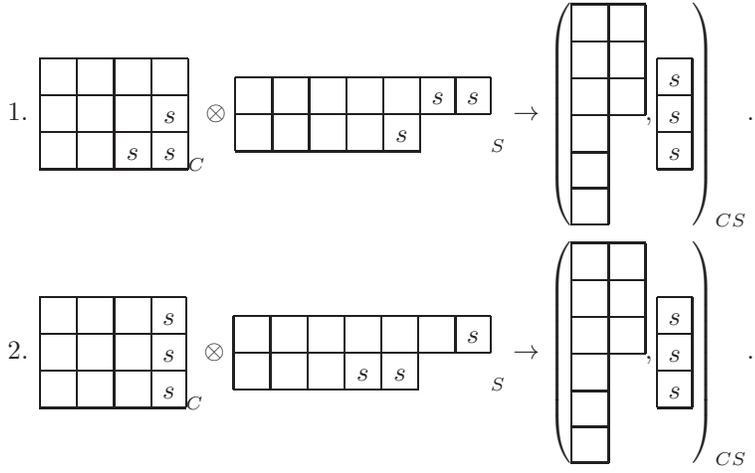
\begin{tabular}{|c|}
  \hline
  $\hspace{0.08cm} s\hspace{0.08cm}$ \\
  \hline
  $\hspace{0.08cm} s\hspace{0.08cm}$ \\
  \hline
  $\hspace{0.08cm} s\hspace{0.08cm}$ \\
  \hline
\end{tabular}
 \right)_{CS}$. \\

2. $\begin{tabular}{|c|c|c|c|}
  \hline
  \quad \quad & \quad \quad & \quad \quad & $\hspace{0.08cm} s\hspace{0.08cm}$ \\
  \hline
  \quad \quad & \quad \quad & \quad \quad & $\hspace{0.08cm} s\hspace{0.08cm}$ \\
  \hline
  \quad \quad & \quad \quad & \quad \quad & $\hspace{0.08cm} s\hspace{0.08cm}$ \\
  \hline
\end{tabular}_C \otimes$
$\begin{tabular}{|c|c|c|c|c|c|c|}
  \hline
  \quad \quad & \quad \quad & \quad \quad & \quad \quad & \quad \quad & \quad \quad & $\hspace{0.08cm} s\hspace{0.08cm}$ \\
  \hline
  \quad \quad & \quad \quad & \quad \quad & $\hspace{0.08cm} s\hspace{0.08cm}$ & $\hspace{0.08cm} s\hspace{0.08cm}$ \\
  \cline{1-5}
\end{tabular}_S \rightarrow$
$\left(\begin{tabular}{|c|c|}
  \hline
  \quad \quad & \quad \quad \\
  \hline
  \quad \quad & \quad \quad \\
  \hline
  \quad \quad & \quad \quad \\
  \hline
  \quad \quad \\
  \cline{1-1}
  \quad \quad \\
  \cline{1-1}
  \quad \quad \\
  \cline{1-1}
\end{tabular},
\begin{tabular}{|c|}
  \hline
  $\hspace{0.08cm} s\hspace{0.08cm}$ \\
  \hline
  $\hspace{0.08cm} s\hspace{0.08cm}$ \\
  \hline
  $\hspace{0.08cm} s\hspace{0.08cm}$ \\
  \hline
\end{tabular}
 \right)_{CS}$.

 \section{$q^{9} s^3(I=\frac{3}{2},S=0)$}

1. $\begin{tabular}{|c|c|c|c|}
  \hline
  \quad \quad & \quad \quad & \quad \quad & \quad \quad \\
  \hline
  \quad \quad & \quad \quad & \quad \quad & $\hspace{0.08cm} s\hspace{0.08cm}$ \\
  \hline
  \quad \quad & \quad \quad & $\hspace{0.08cm} s\hspace{0.08cm}$ & $\hspace{0.08cm} s\hspace{0.08cm}$ \\
  \hline
\end{tabular}_C \otimes$
$\begin{tabular}{|c|c|c|c|c|c|}
  \hline
  \quad \quad & \quad \quad & \quad \quad & \quad \quad & \quad \quad & $\hspace{0.08cm} s\hspace{0.08cm}$  \\
  \hline
  \quad \quad & \quad \quad & \quad \quad & \quad \quad   & $\hspace{0.08cm} s\hspace{0.08cm}$ & $\hspace{0.08cm} s\hspace{0.08cm}$ \\
  \hline
\end{tabular}_S \rightarrow$
$\left(\begin{tabular}{|c|c|}
  \hline
  \quad \quad & \quad \quad \\
  \hline
  \quad \quad & \quad \quad \\
  \hline
  \quad \quad & \quad \quad \\
  \hline
  \quad \quad \\
  \cline{1-1}
  \quad \quad \\
  \cline{1-1}
  \quad \quad \\
  \cline{1-1}
\end{tabular},
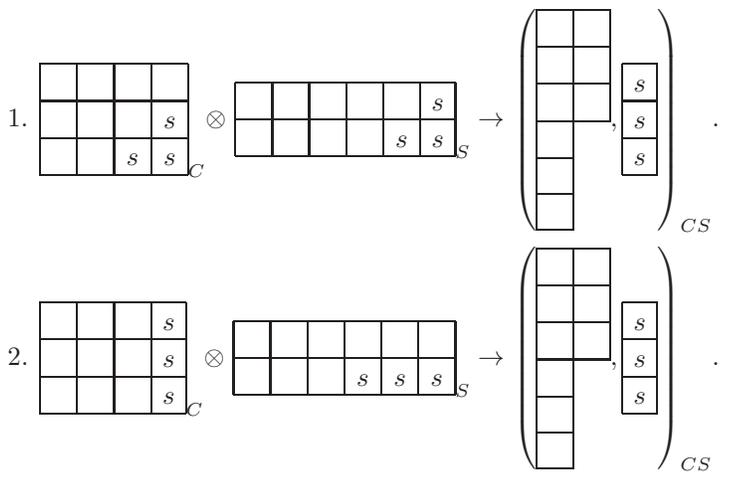
\begin{tabular}{|c|}
  \hline
  $\hspace{0.08cm} s\hspace{0.08cm}$ \\
  \hline
  $\hspace{0.08cm} s\hspace{0.08cm}$ \\
  \hline
  $\hspace{0.08cm} s\hspace{0.08cm}$ \\
  \hline
\end{tabular}
 \right)_{CS}$. \\

2. $\begin{tabular}{|c|c|c|c|}
  \hline
  \quad \quad & \quad \quad & \quad \quad & $\hspace{0.08cm} s\hspace{0.08cm}$ \\
  \hline
  \quad \quad & \quad \quad & \quad \quad & $\hspace{0.08cm} s\hspace{0.08cm}$ \\
  \hline
  \quad \quad & \quad \quad & \quad \quad & $\hspace{0.08cm} s\hspace{0.08cm}$ \\
  \hline
\end{tabular}_C \otimes$
$\begin{tabular}{|c|c|c|c|c|c|}
  \hline
  \quad \quad & \quad \quad & \quad \quad & \quad \quad & \quad \quad & \quad \quad \\
  \hline
  \quad \quad & \quad \quad & \quad \quad & $\hspace{0.08cm} s\hspace{0.08cm}$ & $\hspace{0.08cm} s\hspace{0.08cm}$ & $\hspace{0.08cm} s\hspace{0.08cm}$ \\
  \hline
\end{tabular}_S \rightarrow$
$\left(\begin{tabular}{|c|c|}
  \hline
  \quad \quad & \quad \quad \\
  \hline
  \quad \quad & \quad \quad \\
  \hline
  \quad \quad & \quad \quad \\
  \hline
  \quad \quad \\
  \cline{1-1}
  \quad \quad \\
  \cline{1-1}
  \quad \quad \\
  \cline{1-1}
\end{tabular},
\begin{tabular}{|c|}
  \hline
  $\hspace{0.08cm} s\hspace{0.08cm}$ \\
  \hline
  $\hspace{0.08cm} s\hspace{0.08cm}$ \\
  \hline
  $\hspace{0.08cm} s\hspace{0.08cm}$ \\
  \hline
\end{tabular}
 \right)_{CS}$.

\section{$q^{9} s^3(I=\frac{1}{2},S=2)$}

1. $\begin{tabular}{|c|c|c|c|}
  \hline
  \quad \quad & \quad \quad & \quad \quad & \quad \quad \\
  \hline
  \quad \quad & \quad \quad & \quad \quad & $\hspace{0.08cm} s\hspace{0.08cm}$ \\
  \hline
  \quad \quad & \quad \quad & $\hspace{0.08cm} s\hspace{0.08cm}$ & $\hspace{0.08cm} s\hspace{0.08cm}$ \\
  \hline
\end{tabular}_C \otimes$
$\begin{tabular}{|c|c|c|c|c|c|c|c|}
  \hline
  \quad \quad & \quad \quad & \quad \quad & \quad \quad & \quad \quad & \quad \quad & $\hspace{0.08cm} s\hspace{0.08cm}$ & $\hspace{0.08cm} s\hspace{0.08cm}$  \\
  \hline
  \quad \quad & \quad \quad & \quad \quad &  $\hspace{0.08cm} s\hspace{0.08cm}$  \\
  \cline{1-4}
\end{tabular}_S \rightarrow$
$\left(\begin{tabular}{|c|c|}
  \hline
  \quad \quad & \quad \quad \\
  \hline
  \quad \quad & \quad \quad \\
  \hline
  \quad \quad & \quad \quad \\
  \hline
  \quad \quad & \quad \quad \\
  \hline
  \quad \quad \\
  \cline{1-1}
\end{tabular},
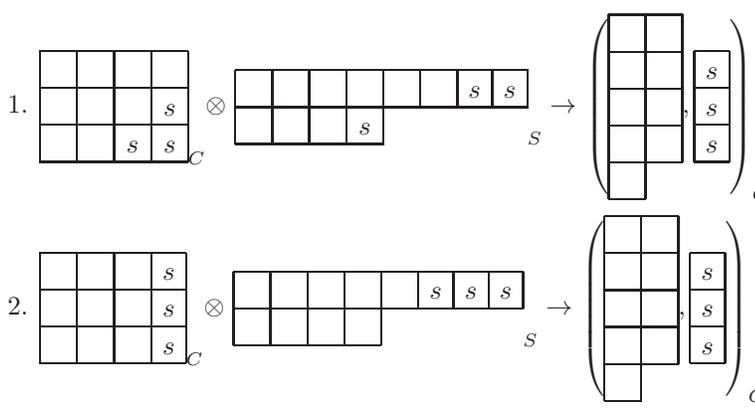
\begin{tabular}{|c|}
  \hline
  $\hspace{0.08cm} s\hspace{0.08cm}$ \\
  \hline
  $\hspace{0.08cm} s\hspace{0.08cm}$ \\
  \hline
  $\hspace{0.08cm} s\hspace{0.08cm}$ \\
  \hline
\end{tabular}
 \right)_{CS}$. \\

2. $\begin{tabular}{|c|c|c|c|}
  \hline
  \quad \quad & \quad \quad & \quad \quad & $\hspace{0.08cm} s\hspace{0.08cm}$ \\
  \hline
  \quad \quad & \quad \quad & \quad \quad & $\hspace{0.08cm} s\hspace{0.08cm}$ \\
  \hline
  \quad \quad & \quad \quad & \quad \quad & $\hspace{0.08cm} s\hspace{0.08cm}$ \\
  \hline
\end{tabular}_C \otimes$
$\begin{tabular}{|c|c|c|c|c|c|c|c|}
  \hline
  \quad \quad & \quad \quad & \quad \quad & \quad \quad & \quad \quad & $\hspace{0.08cm} s\hspace{0.08cm}$ & $\hspace{0.08cm} s\hspace{0.08cm}$ & $\hspace{0.08cm} s\hspace{0.08cm}$ \\
  \hline
  \quad \quad & \quad \quad & \quad \quad & \quad \quad \\
  \cline{1-4}
\end{tabular}_S \rightarrow$
$\left(\begin{tabular}{|c|c|}
  \hline
  \quad \quad & \quad \quad \\
  \hline
  \quad \quad & \quad \quad \\
  \hline
  \quad \quad & \quad \quad \\
  \hline
  \quad \quad & \quad \quad \\
  \hline
  \quad \quad \\
  \cline{1-1}
\end{tabular},
\begin{tabular}{|c|}
  \hline
  $\hspace{0.08cm} s\hspace{0.08cm}$ \\
  \hline
  $\hspace{0.08cm} s\hspace{0.08cm}$ \\
  \hline
  $\hspace{0.08cm} s\hspace{0.08cm}$ \\
  \hline
\end{tabular}
 \right)_{CS}$.

\section{$q^{9} s^3(I=\frac{1}{2},S=1)$}

1. $\begin{tabular}{|c|c|c|c|}
  \hline
  \quad \quad & \quad \quad & \quad \quad & \quad \quad \\
  \hline
  \quad \quad & \quad \quad & \quad \quad & $\hspace{0.08cm} s\hspace{0.08cm}$ \\
  \hline
  \quad \quad & \quad \quad & $\hspace{0.08cm} s\hspace{0.08cm}$ & $\hspace{0.08cm} s\hspace{0.08cm}$ \\
  \hline
\end{tabular}_C \otimes$
$\begin{tabular}{|c|c|c|c|c|c|c|}
  \hline
  \quad \quad & \quad \quad & \quad \quad & \quad \quad & \quad \quad & \quad \quad & $\hspace{0.08cm} s\hspace{0.08cm}$  \\
  \hline
  \quad \quad & \quad \quad & \quad \quad &  $\hspace{0.08cm} s\hspace{0.08cm}$ & $\hspace{0.08cm} s\hspace{0.08cm}$  \\
  \cline{1-5}
\end{tabular}_S \rightarrow$
$\left(\begin{tabular}{|c|c|}
  \hline
  \quad \quad & \quad \quad \\
  \hline
  \quad \quad & \quad \quad \\
  \hline
  \quad \quad & \quad \quad \\
  \hline
  \quad \quad & \quad \quad \\
  \hline
  \quad \quad \\
  \cline{1-1}
\end{tabular},
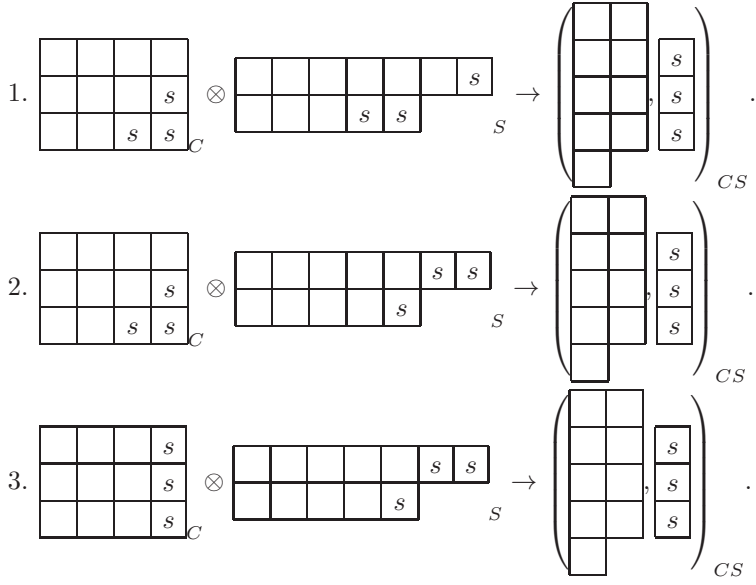
\begin{tabular}{|c|}
  \hline
  $\hspace{0.08cm} s\hspace{0.08cm}$ \\
  \hline
  $\hspace{0.08cm} s\hspace{0.08cm}$ \\
  \hline
  $\hspace{0.08cm} s\hspace{0.08cm}$ \\
  \hline
\end{tabular}
 \right)_{CS}$. \\

2. $\begin{tabular}{|c|c|c|c|}
  \hline
  \quad \quad & \quad \quad & \quad \quad & \quad \quad \\
  \hline
  \quad \quad & \quad \quad & \quad \quad & $\hspace{0.08cm} s\hspace{0.08cm}$ \\
  \hline
  \quad \quad & \quad \quad & $\hspace{0.08cm} s\hspace{0.08cm}$ & $\hspace{0.08cm} s\hspace{0.08cm}$ \\
  \hline
\end{tabular}_C \otimes$
$\begin{tabular}{|c|c|c|c|c|c|c|}
  \hline
  \quad \quad & \quad \quad & \quad \quad & \quad \quad & \quad \quad & $\hspace{0.08cm} s\hspace{0.08cm}$ & $\hspace{0.08cm} s\hspace{0.08cm}$  \\
  \hline
  \quad \quad & \quad \quad & \quad \quad &  \quad \quad & $\hspace{0.08cm} s\hspace{0.08cm}$  \\
  \cline{1-5}
\end{tabular}_S \rightarrow$
$\left(\begin{tabular}{|c|c|}
  \hline
  \quad \quad & \quad \quad \\
  \hline
  \quad \quad & \quad \quad \\
  \hline
  \quad \quad & \quad \quad \\
  \hline
  \quad \quad & \quad \quad \\
  \hline
  \quad \quad \\
  \cline{1-1}
\end{tabular},
\begin{tabular}{|c|}
  \hline
  $\hspace{0.08cm} s\hspace{0.08cm}$ \\
  \hline
  $\hspace{0.08cm} s\hspace{0.08cm}$ \\
  \hline
  $\hspace{0.08cm} s\hspace{0.08cm}$ \\
  \hline
\end{tabular}
 \right)_{CS}$. \\

3. $\begin{tabular}{|c|c|c|c|}
  \hline
  \quad \quad & \quad \quad & \quad \quad & $\hspace{0.08cm} s\hspace{0.08cm}$ \\
  \hline
  \quad \quad & \quad \quad & \quad \quad & $\hspace{0.08cm} s\hspace{0.08cm}$ \\
  \hline
  \quad \quad & \quad \quad & \quad \quad & $\hspace{0.08cm} s\hspace{0.08cm}$ \\
  \hline
\end{tabular}_C \otimes$
$\begin{tabular}{|c|c|c|c|c|c|c|}
  \hline
  \quad \quad & \quad \quad & \quad \quad & \quad \quad & \quad \quad & $\hspace{0.08cm} s\hspace{0.08cm}$ & $\hspace{0.08cm} s\hspace{0.08cm}$ \\
  \hline
  \quad \quad & \quad \quad & \quad \quad & \quad \quad & $\hspace{0.08cm} s\hspace{0.08cm}$ \\
  \cline{1-5}
\end{tabular}_S \rightarrow$
$\left(\begin{tabular}{|c|c|}
  \hline
  \quad \quad & \quad \quad \\
  \hline
  \quad \quad & \quad \quad \\
  \hline
  \quad \quad & \quad \quad \\
  \hline
  \quad \quad & \quad \quad \\
  \hline
  \quad \quad \\
  \cline{1-1}
\end{tabular},
\begin{tabular}{|c|}
  \hline
  $\hspace{0.08cm} s\hspace{0.08cm}$ \\
  \hline
  $\hspace{0.08cm} s\hspace{0.08cm}$ \\
  \hline
  $\hspace{0.08cm} s\hspace{0.08cm}$ \\
  \hline
\end{tabular}
 \right)_{CS}$.

 \section{$q^{8} s^4(I=2,S=0)$}

1. $\begin{tabular}{|c|c|c|c|}
  \hline
  \quad \quad & \quad \quad & \quad \quad & \quad \quad \\
  \hline
  \quad \quad & \quad \quad & $\hspace{0.08cm} s\hspace{0.08cm}$ & $\hspace{0.08cm} s\hspace{0.08cm}$ \\
  \hline
  \quad \quad & \quad \quad & $\hspace{0.08cm} s\hspace{0.08cm}$ & $\hspace{0.08cm} s\hspace{0.08cm}$ \\
  \hline
\end{tabular}_C \otimes$
$\begin{tabular}{|c|c|c|c|c|c|}
  \hline
  \quad \quad & \quad \quad & \quad \quad & \quad \quad &  $\hspace{0.08cm} s\hspace{0.08cm}$ &  $\hspace{0.08cm} s\hspace{0.08cm}$ \\
  \hline
  \quad \quad & \quad \quad & \quad \quad & \quad \quad &  $\hspace{0.08cm} s\hspace{0.08cm}$ &  $\hspace{0.08cm} s\hspace{0.08cm}$ \\
  \hline
\end{tabular}_S \rightarrow$
$\left(\begin{tabular}{|c|c|}
  \hline
  \quad \quad & \quad \quad \\
  \hline
  \quad \quad & \quad \quad \\
  \hline
  \quad \quad \\
  \cline{1-1}
  \quad \quad \\
  \cline{1-1}
  \quad \quad \\
  \cline{1-1}
  \quad \quad \\
  \cline{1-1}
\end{tabular},
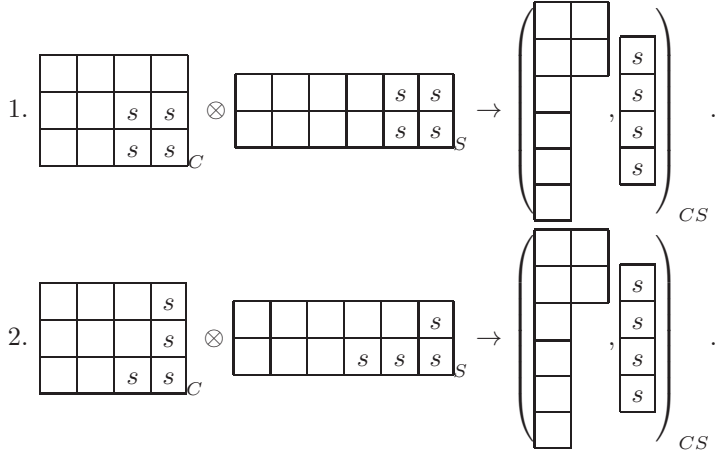
\begin{tabular}{|c|}
  \hline
  $\hspace{0.08cm} s\hspace{0.08cm}$ \\
  \hline
  $\hspace{0.08cm} s\hspace{0.08cm}$ \\
  \hline
  $\hspace{0.08cm} s\hspace{0.08cm}$ \\
  \hline
  $\hspace{0.08cm} s\hspace{0.08cm}$ \\
  \hline
\end{tabular}
 \right)_{CS}$. \\

2. $\begin{tabular}{|c|c|c|c|}
  \hline
  \quad \quad & \quad \quad & \quad \quad & $\hspace{0.08cm} s\hspace{0.08cm}$ \\
  \hline
  \quad \quad & \quad \quad & \quad \quad & $\hspace{0.08cm} s\hspace{0.08cm}$ \\
  \hline
  \quad \quad & \quad \quad & $\hspace{0.08cm} s\hspace{0.08cm}$ & $\hspace{0.08cm} s\hspace{0.08cm}$ \\
  \hline
\end{tabular}_C \otimes$
$\begin{tabular}{|c|c|c|c|c|c|}
  \hline
  \quad \quad & \quad \quad & \quad \quad & \quad \quad &  \quad \quad &  $\hspace{0.08cm} s\hspace{0.08cm}$ \\
  \hline
  \quad \quad & \quad \quad & \quad \quad & $\hspace{0.08cm} s\hspace{0.08cm}$ &  $\hspace{0.08cm} s\hspace{0.08cm}$ &  $\hspace{0.08cm} s\hspace{0.08cm}$ \\
  \hline
\end{tabular}_S \rightarrow$
$\left(\begin{tabular}{|c|c|}
  \hline
  \quad \quad & \quad \quad \\
  \hline
  \quad \quad & \quad \quad \\
  \hline
  \quad \quad \\
  \cline{1-1}
  \quad \quad \\
  \cline{1-1}
  \quad \quad \\
  \cline{1-1}
  \quad \quad \\
  \cline{1-1}
\end{tabular},
\begin{tabular}{|c|}
  \hline
  $\hspace{0.08cm} s\hspace{0.08cm}$ \\
  \hline
  $\hspace{0.08cm} s\hspace{0.08cm}$ \\
  \hline
  $\hspace{0.08cm} s\hspace{0.08cm}$ \\
  \hline
  $\hspace{0.08cm} s\hspace{0.08cm}$ \\
  \hline
\end{tabular}
 \right)_{CS}$. \\

\section{$q^{8} s^4(I=1,S=2)$}

1. $\begin{tabular}{|c|c|c|c|}
  \hline
  \quad \quad & \quad \quad & \quad \quad & $\hspace{0.08cm} s\hspace{0.08cm}$ \\
  \hline
  \quad \quad & \quad \quad & \quad \quad & $\hspace{0.08cm} s\hspace{0.08cm}$ \\
  \hline
  \quad \quad & \quad \quad & $\hspace{0.08cm} s\hspace{0.08cm}$ & $\hspace{0.08cm} s\hspace{0.08cm}$ \\
  \hline
\end{tabular}_C \otimes$
$\begin{tabular}{|c|c|c|c|c|c|c|c|}
  \hline
  \quad \quad & \quad \quad & \quad \quad & \quad \quad & \quad \quad & \quad \quad & $\hspace{0.08cm} s\hspace{0.08cm}$ &  $\hspace{0.08cm} s\hspace{0.08cm}$ \\
  \hline
  \quad \quad & \quad \quad & $\hspace{0.08cm} s\hspace{0.08cm}$ &  $\hspace{0.08cm} s\hspace{0.08cm}$ \\
  \cline{1-4}
\end{tabular}_S \rightarrow$
$\left(\begin{tabular}{|c|c|}
  \hline
  \quad \quad & \quad \quad \\
  \hline
  \quad \quad & \quad \quad \\
  \hline
  \quad \quad & \quad \quad \\
  \hline
  \quad \quad \\
  \cline{1-1}
  \quad \quad \\
  \cline{1-1}
\end{tabular},
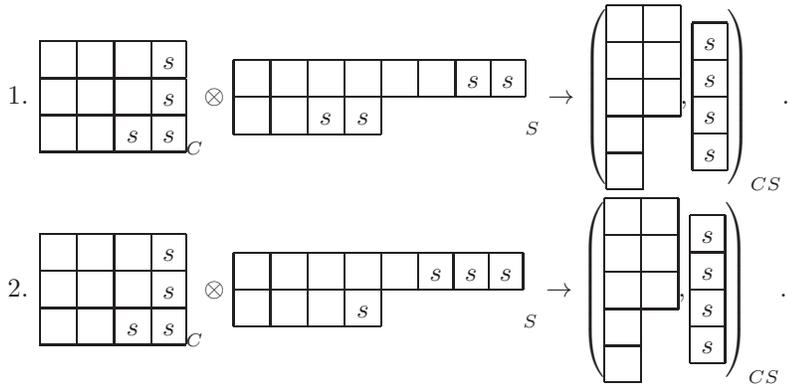
\begin{tabular}{|c|}
  \hline
  $\hspace{0.08cm} s\hspace{0.08cm}$ \\
  \hline
  $\hspace{0.08cm} s\hspace{0.08cm}$ \\
  \hline
  $\hspace{0.08cm} s\hspace{0.08cm}$ \\
  \hline
  $\hspace{0.08cm} s\hspace{0.08cm}$ \\
  \hline
\end{tabular}
 \right)_{CS}$. \\

2. $\begin{tabular}{|c|c|c|c|}
  \hline
  \quad \quad & \quad \quad & \quad \quad & $\hspace{0.08cm} s\hspace{0.08cm}$ \\
  \hline
  \quad \quad & \quad \quad & \quad \quad & $\hspace{0.08cm} s\hspace{0.08cm}$ \\
  \hline
  \quad \quad & \quad \quad & $\hspace{0.08cm} s\hspace{0.08cm}$ & $\hspace{0.08cm} s\hspace{0.08cm}$ \\
  \hline
\end{tabular}_C \otimes$
$\begin{tabular}{|c|c|c|c|c|c|c|c|}
  \hline
  \quad \quad & \quad \quad & \quad \quad & \quad \quad & \quad \quad & $\hspace{0.08cm} s\hspace{0.08cm}$ & $\hspace{0.08cm} s\hspace{0.08cm}$ &  $\hspace{0.08cm} s\hspace{0.08cm}$ \\
  \hline
  \quad \quad & \quad \quad & \quad \quad &  $\hspace{0.08cm} s\hspace{0.08cm}$ \\
  \cline{1-4}
\end{tabular}_S \rightarrow$
$\left(\begin{tabular}{|c|c|}
  \hline
  \quad \quad & \quad \quad \\
  \hline
  \quad \quad & \quad \quad \\
  \hline
  \quad \quad & \quad \quad \\
  \hline
  \quad \quad \\
  \cline{1-1}
  \quad \quad \\
  \cline{1-1}
\end{tabular},
\begin{tabular}{|c|}
  \hline
  $\hspace{0.08cm} s\hspace{0.08cm}$ \\
  \hline
  $\hspace{0.08cm} s\hspace{0.08cm}$ \\
  \hline
  $\hspace{0.08cm} s\hspace{0.08cm}$ \\
  \hline
  $\hspace{0.08cm} s\hspace{0.08cm}$ \\
  \hline
\end{tabular}
 \right)_{CS}$. \\

\section{$q^{8} s^4(I=1,S=1)$}

1. $\begin{tabular}{|c|c|c|c|}
  \hline
  \quad \quad & \quad \quad & \quad \quad & \quad \quad \\
  \hline
  \quad \quad & \quad \quad & $\hspace{0.08cm} s\hspace{0.08cm}$ & $\hspace{0.08cm} s\hspace{0.08cm}$ \\
  \hline
  \quad \quad & \quad \quad & $\hspace{0.08cm} s\hspace{0.08cm}$ & $\hspace{0.08cm} s\hspace{0.08cm}$ \\
  \hline
\end{tabular}_C \otimes$
$\begin{tabular}{|c|c|c|c|c|c|c|}
  \hline
  \quad \quad & \quad \quad & \quad \quad & \quad \quad & \quad \quad & $\hspace{0.08cm} s\hspace{0.08cm}$ & $\hspace{0.08cm} s\hspace{0.08cm}$  \\
  \hline
  \quad \quad & \quad \quad & \quad \quad &  $\hspace{0.08cm} s\hspace{0.08cm}$ &  $\hspace{0.08cm} s\hspace{0.08cm}$ \\
  \cline{1-5}
\end{tabular}_S \rightarrow$
$\left(\begin{tabular}{|c|c|}
  \hline
  \quad \quad & \quad \quad \\
  \hline
  \quad \quad & \quad \quad \\
  \hline
  \quad \quad & \quad \quad \\
  \hline
  \quad \quad \\
  \cline{1-1}
  \quad \quad \\
  \cline{1-1}
\end{tabular},
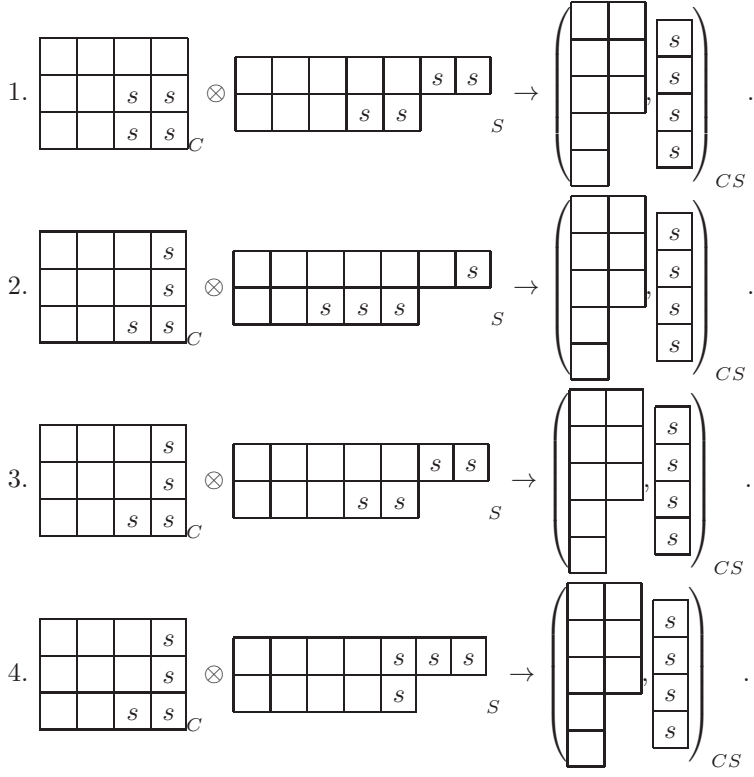
\begin{tabular}{|c|}
  \hline
  $\hspace{0.08cm} s\hspace{0.08cm}$ \\
  \hline
  $\hspace{0.08cm} s\hspace{0.08cm}$ \\
  \hline
  $\hspace{0.08cm} s\hspace{0.08cm}$ \\
  \hline
  $\hspace{0.08cm} s\hspace{0.08cm}$ \\
  \hline
\end{tabular}
 \right)_{CS}$. \\

2. $\begin{tabular}{|c|c|c|c|}
  \hline
  \quad \quad & \quad \quad & \quad \quad & $\hspace{0.08cm} s\hspace{0.08cm}$ \\
  \hline
  \quad \quad & \quad \quad & \quad \quad & $\hspace{0.08cm} s\hspace{0.08cm}$ \\
  \hline
  \quad \quad & \quad \quad & $\hspace{0.08cm} s\hspace{0.08cm}$ & $\hspace{0.08cm} s\hspace{0.08cm}$ \\
  \hline
\end{tabular}_C \otimes$
$\begin{tabular}{|c|c|c|c|c|c|c|c|}
  \hline
  \quad \quad & \quad \quad & \quad \quad & \quad \quad & \quad \quad & \quad \quad & $\hspace{0.08cm} s\hspace{0.08cm}$ \\
  \hline
  \quad \quad & \quad \quad & $\hspace{0.08cm} s\hspace{0.08cm}$ &  $\hspace{0.08cm} s\hspace{0.08cm}$ &  $\hspace{0.08cm} s\hspace{0.08cm}$ \\
  \cline{1-5}
\end{tabular}_S \rightarrow$
$\left(\begin{tabular}{|c|c|}
  \hline
  \quad \quad & \quad \quad \\
  \hline
  \quad \quad & \quad \quad \\
  \hline
  \quad \quad & \quad \quad \\
  \hline
  \quad \quad \\
  \cline{1-1}
  \quad \quad \\
  \cline{1-1}
\end{tabular},
\begin{tabular}{|c|}
  \hline
  $\hspace{0.08cm} s\hspace{0.08cm}$ \\
  \hline
  $\hspace{0.08cm} s\hspace{0.08cm}$ \\
  \hline
  $\hspace{0.08cm} s\hspace{0.08cm}$ \\
  \hline
  $\hspace{0.08cm} s\hspace{0.08cm}$ \\
  \hline
\end{tabular}
 \right)_{CS}$. \\

3. $\begin{tabular}{|c|c|c|c|}
  \hline
  \quad \quad & \quad \quad & \quad \quad & $\hspace{0.08cm} s\hspace{0.08cm}$ \\
  \hline
  \quad \quad & \quad \quad & \quad \quad & $\hspace{0.08cm} s\hspace{0.08cm}$ \\
  \hline
  \quad \quad & \quad \quad & $\hspace{0.08cm} s\hspace{0.08cm}$ & $\hspace{0.08cm} s\hspace{0.08cm}$ \\
  \hline
\end{tabular}_C \otimes$
$\begin{tabular}{|c|c|c|c|c|c|c|c|}
  \hline
  \quad \quad & \quad \quad & \quad \quad & \quad \quad & \quad \quad & $\hspace{0.08cm} s\hspace{0.08cm}$ & $\hspace{0.08cm} s\hspace{0.08cm}$ \\
  \hline
  \quad \quad & \quad \quad & \quad \quad &  $\hspace{0.08cm} s\hspace{0.08cm}$ &  $\hspace{0.08cm} s\hspace{0.08cm}$ \\
  \cline{1-5}
\end{tabular}_S \rightarrow$
$\left(\begin{tabular}{|c|c|}
  \hline
  \quad \quad & \quad \quad \\
  \hline
  \quad \quad & \quad \quad \\
  \hline
  \quad \quad & \quad \quad \\
  \hline
  \quad \quad \\
  \cline{1-1}
  \quad \quad \\
  \cline{1-1}
\end{tabular},
\begin{tabular}{|c|}
  \hline
  $\hspace{0.08cm} s\hspace{0.08cm}$ \\
  \hline
  $\hspace{0.08cm} s\hspace{0.08cm}$ \\
  \hline
  $\hspace{0.08cm} s\hspace{0.08cm}$ \\
  \hline
  $\hspace{0.08cm} s\hspace{0.08cm}$ \\
  \hline
\end{tabular}
 \right)_{CS}$. \\

4. $\begin{tabular}{|c|c|c|c|}
  \hline
  \quad \quad & \quad \quad & \quad \quad & $\hspace{0.08cm} s\hspace{0.08cm}$ \\
  \hline
  \quad \quad & \quad \quad & \quad \quad & $\hspace{0.08cm} s\hspace{0.08cm}$ \\
  \hline
  \quad \quad & \quad \quad & $\hspace{0.08cm} s\hspace{0.08cm}$ & $\hspace{0.08cm} s\hspace{0.08cm}$ \\
  \hline
\end{tabular}_C \otimes$
$\begin{tabular}{|c|c|c|c|c|c|c|c|}
  \hline
  \quad \quad & \quad \quad & \quad \quad & \quad \quad & $\hspace{0.08cm} s\hspace{0.08cm}$ & $\hspace{0.08cm} s\hspace{0.08cm}$ & $\hspace{0.08cm} s\hspace{0.08cm}$ \\
  \hline
  \quad \quad & \quad \quad & \quad \quad &  \quad \quad & $\hspace{0.08cm} s\hspace{0.08cm}$ \\
  \cline{1-5}
\end{tabular}_S \rightarrow$
$\left(\begin{tabular}{|c|c|}
  \hline
  \quad \quad & \quad \quad \\
  \hline
  \quad \quad & \quad \quad \\
  \hline
  \quad \quad & \quad \quad \\
  \hline
  \quad \quad \\
  \cline{1-1}
  \quad \quad \\
  \cline{1-1}
\end{tabular},
\begin{tabular}{|c|}
  \hline
  $\hspace{0.08cm} s\hspace{0.08cm}$ \\
  \hline
  $\hspace{0.08cm} s\hspace{0.08cm}$ \\
  \hline
  $\hspace{0.08cm} s\hspace{0.08cm}$ \\
  \hline
  $\hspace{0.08cm} s\hspace{0.08cm}$ \\
  \hline
\end{tabular}
 \right)_{CS}$. \\

\section{$q^{8} s^4(I=0,S=2)$}

1. $\begin{tabular}{|c|c|c|c|}
  \hline
  \quad \quad & \quad \quad & \quad \quad & \quad \quad \\
  \hline
  \quad \quad & \quad \quad & $\hspace{0.08cm} s\hspace{0.08cm}$ & $\hspace{0.08cm} s\hspace{0.08cm}$ \\
  \hline
  \quad \quad & \quad \quad & $\hspace{0.08cm} s\hspace{0.08cm}$ & $\hspace{0.08cm} s\hspace{0.08cm}$ \\
  \hline
\end{tabular}_C \otimes$
$\begin{tabular}{|c|c|c|c|c|c|c|c|}
  \hline
  \quad \quad & \quad \quad & \quad \quad & \quad \quad & \quad \quad & \quad \quad & $\hspace{0.08cm} s\hspace{0.08cm}$ &  $\hspace{0.08cm} s\hspace{0.08cm}$ \\
  \hline
  \quad \quad & \quad \quad & $\hspace{0.08cm} s\hspace{0.08cm}$ &  $\hspace{0.08cm} s\hspace{0.08cm}$ \\
  \cline{1-4}
\end{tabular}_S \rightarrow$
$\left(\begin{tabular}{|c|c|}
  \hline
  \quad \quad & \quad \quad \\
  \hline
  \quad \quad & \quad \quad \\
  \hline
  \quad \quad & \quad \quad \\
  \hline
  \quad \quad & \quad \quad \\
  \hline
\end{tabular},
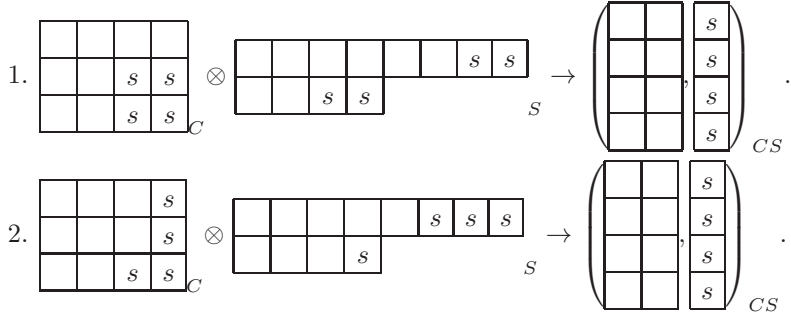
\begin{tabular}{|c|}
  \hline
  $\hspace{0.08cm} s\hspace{0.08cm}$ \\
  \hline
  $\hspace{0.08cm} s\hspace{0.08cm}$ \\
  \hline
  $\hspace{0.08cm} s\hspace{0.08cm}$ \\
  \hline
  $\hspace{0.08cm} s\hspace{0.08cm}$ \\
  \hline
\end{tabular}
 \right)_{CS}$. \\

2. $\begin{tabular}{|c|c|c|c|}
  \hline
  \quad \quad & \quad \quad & \quad \quad & $\hspace{0.08cm} s\hspace{0.08cm}$ \\
  \hline
  \quad \quad & \quad \quad & \quad \quad & $\hspace{0.08cm} s\hspace{0.08cm}$ \\
  \hline
  \quad \quad & \quad \quad & $\hspace{0.08cm} s\hspace{0.08cm}$ & $\hspace{0.08cm} s\hspace{0.08cm}$ \\
  \hline
\end{tabular}_C \otimes$
$\begin{tabular}{|c|c|c|c|c|c|c|c|}
  \hline
  \quad \quad & \quad \quad & \quad \quad & \quad \quad & \quad \quad & $\hspace{0.08cm} s\hspace{0.08cm}$ & $\hspace{0.08cm} s\hspace{0.08cm}$ &  $\hspace{0.08cm} s\hspace{0.08cm}$ \\
  \hline
  \quad \quad & \quad \quad & \quad \quad &  $\hspace{0.08cm} s\hspace{0.08cm}$ \\
  \cline{1-4}
\end{tabular}_S \rightarrow$
$\left(\begin{tabular}{|c|c|}
  \hline
  \quad \quad & \quad \quad \\
  \hline
  \quad \quad & \quad \quad \\
  \hline
  \quad \quad & \quad \quad \\
  \hline
  \quad \quad & \quad \quad \\
  \hline
\end{tabular},
\begin{tabular}{|c|}
  \hline
  $\hspace{0.08cm} s\hspace{0.08cm}$ \\
  \hline
  $\hspace{0.08cm} s\hspace{0.08cm}$ \\
  \hline
  $\hspace{0.08cm} s\hspace{0.08cm}$ \\
  \hline
  $\hspace{0.08cm} s\hspace{0.08cm}$ \\
  \hline
\end{tabular}
 \right)_{CS}$. \\

\section{$q^{8} s^4(I=0,S=0)$}

1. $\begin{tabular}{|c|c|c|c|}
  \hline
  \quad \quad & \quad \quad & \quad \quad & \quad \quad \\
  \hline
  \quad \quad & \quad \quad & $\hspace{0.08cm} s\hspace{0.08cm}$ & $\hspace{0.08cm} s\hspace{0.08cm}$ \\
  \hline
  \quad \quad & \quad \quad & $\hspace{0.08cm} s\hspace{0.08cm}$ & $\hspace{0.08cm} s\hspace{0.08cm}$ \\
  \hline
\end{tabular}_C \otimes$
$\begin{tabular}{|c|c|c|c|c|c|}
  \hline
  \quad \quad & \quad \quad & \quad \quad & \quad \quad & $\hspace{0.08cm} s\hspace{0.08cm}$ &  $\hspace{0.08cm} s\hspace{0.08cm}$ \\
  \hline
  \quad \quad & \quad \quad & \quad \quad & \quad \quad & $\hspace{0.08cm} s\hspace{0.08cm}$ &  $\hspace{0.08cm} s\hspace{0.08cm}$ \\
  \hline
\end{tabular}_S \rightarrow$
$\left(\begin{tabular}{|c|c|}
  \hline
  \quad \quad & \quad \quad \\
  \hline
  \quad \quad & \quad \quad \\
  \hline
  \quad \quad & \quad \quad \\
  \hline
  \quad \quad & \quad \quad \\
  \hline
\end{tabular},
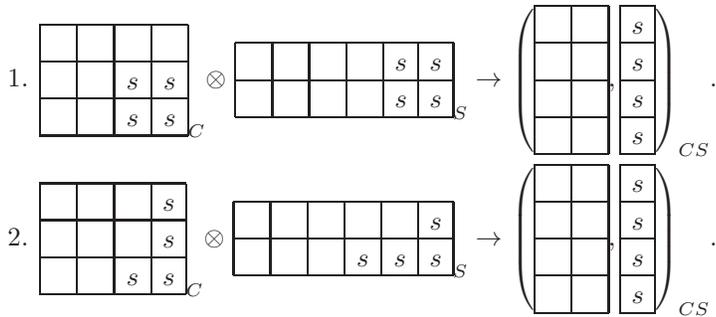
\begin{tabular}{|c|}
  \hline
  $\hspace{0.08cm} s\hspace{0.08cm}$ \\
  \hline
  $\hspace{0.08cm} s\hspace{0.08cm}$ \\
  \hline
  $\hspace{0.08cm} s\hspace{0.08cm}$ \\
  \hline
  $\hspace{0.08cm} s\hspace{0.08cm}$ \\
  \hline
\end{tabular}
 \right)_{CS}$. \\

2. $\begin{tabular}{|c|c|c|c|}
  \hline
  \quad \quad & \quad \quad & \quad \quad & $\hspace{0.08cm} s\hspace{0.08cm}$ \\
  \hline
  \quad \quad & \quad \quad & \quad \quad & $\hspace{0.08cm} s\hspace{0.08cm}$ \\
  \hline
  \quad \quad & \quad \quad & $\hspace{0.08cm} s\hspace{0.08cm}$ & $\hspace{0.08cm} s\hspace{0.08cm}$ \\
  \hline
\end{tabular}_C \otimes$
$\begin{tabular}{|c|c|c|c|c|c|}
  \hline
  \quad \quad & \quad \quad & \quad \quad & \quad \quad & \quad \quad & $\hspace{0.08cm} s\hspace{0.08cm}$  \\
  \hline
  \quad \quad & \quad \quad & \quad \quad &  $\hspace{0.08cm} s\hspace{0.08cm}$ & $\hspace{0.08cm} s\hspace{0.08cm}$ &  $\hspace{0.08cm} s\hspace{0.08cm}$ \\
  \hline
\end{tabular}_S \rightarrow$
$\left(\begin{tabular}{|c|c|}
  \hline
  \quad \quad & \quad \quad \\
  \hline
  \quad \quad & \quad \quad \\
  \hline
  \quad \quad & \quad \quad \\
  \hline
  \quad \quad & \quad \quad \\
  \hline
\end{tabular},
\begin{tabular}{|c|}
  \hline
  $\hspace{0.08cm} s\hspace{0.08cm}$ \\
  \hline
  $\hspace{0.08cm} s\hspace{0.08cm}$ \\
  \hline
  $\hspace{0.08cm} s\hspace{0.08cm}$ \\
  \hline
  $\hspace{0.08cm} s\hspace{0.08cm}$ \\
  \hline
\end{tabular}
 \right)_{CS}$. \\

\section{$q^7 s^5(I=\frac{3}{2},S=1)$}

1. $\begin{tabular}{|c|c|c|c|}
  \hline
  \quad \quad & \quad \quad & \quad \quad & $\hspace{0.08cm} s\hspace{0.08cm}$ \\
  \hline
  \quad \quad & \quad \quad & $\hspace{0.08cm} s\hspace{0.08cm}$ & $\hspace{0.08cm} s\hspace{0.08cm}$ \\
  \hline
  \quad \quad & \quad \quad & $\hspace{0.08cm} s\hspace{0.08cm}$ & $\hspace{0.08cm} s\hspace{0.08cm}$ \\
  \hline
\end{tabular}_C \otimes$
$\begin{tabular}{|c|c|c|c|c|c|c|}
  \hline
  \quad \quad & \quad \quad & \quad \quad & \quad \quad & \quad \quad & $\hspace{0.08cm} s\hspace{0.08cm}$ &  $\hspace{0.08cm} s\hspace{0.08cm}$ \\
  \hline
  \quad \quad & \quad \quad & $\hspace{0.08cm} s\hspace{0.08cm}$ & $\hspace{0.08cm} s\hspace{0.08cm}$ &  $\hspace{0.08cm} s\hspace{0.08cm}$ \\
  \cline{1-5}
\end{tabular}_S \rightarrow$
$\left(\begin{tabular}{|c|c|}
  \hline
  \quad \quad & \quad \quad \\
  \hline
  \quad \quad & \quad \quad \\
  \hline
  \quad \quad \\
  \cline{1-1}
  \quad \quad \\
  \cline{1-1}
  \quad \quad \\
  \cline{1-1}
\end{tabular},
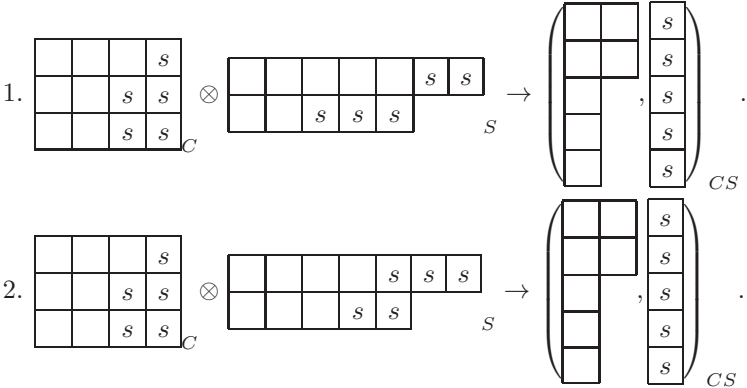
\begin{tabular}{|c|}
  \hline
  $\hspace{0.08cm} s\hspace{0.08cm}$ \\
  \hline
  $\hspace{0.08cm} s\hspace{0.08cm}$ \\
  \hline
  $\hspace{0.08cm} s\hspace{0.08cm}$ \\
  \hline
  $\hspace{0.08cm} s\hspace{0.08cm}$ \\
  \hline
  $\hspace{0.08cm} s\hspace{0.08cm}$ \\
  \hline
\end{tabular}
 \right)_{CS}$. \\

2. $\begin{tabular}{|c|c|c|c|}
  \hline
  \quad \quad & \quad \quad & \quad \quad & $\hspace{0.08cm} s\hspace{0.08cm}$ \\
  \hline
  \quad \quad & \quad \quad & $\hspace{0.08cm} s\hspace{0.08cm}$ & $\hspace{0.08cm} s\hspace{0.08cm}$ \\
  \hline
  \quad \quad & \quad \quad & $\hspace{0.08cm} s\hspace{0.08cm}$ & $\hspace{0.08cm} s\hspace{0.08cm}$ \\
  \hline
\end{tabular}_C \otimes$
$\begin{tabular}{|c|c|c|c|c|c|c|}
  \hline
  \quad \quad & \quad \quad & \quad \quad & \quad \quad & $\hspace{0.08cm} s\hspace{0.08cm}$ & $\hspace{0.08cm} s\hspace{0.08cm}$ &  $\hspace{0.08cm} s\hspace{0.08cm}$ \\
  \hline
  \quad \quad & \quad \quad & \quad \quad & $\hspace{0.08cm} s\hspace{0.08cm}$ &  $\hspace{0.08cm} s\hspace{0.08cm}$ \\
  \cline{1-5}
\end{tabular}_S \rightarrow$
$\left(\begin{tabular}{|c|c|}
  \hline
  \quad \quad & \quad \quad \\
  \hline
  \quad \quad & \quad \quad \\
  \hline
  \quad \quad \\
  \cline{1-1}
  \quad \quad \\
  \cline{1-1}
  \quad \quad \\
  \cline{1-1}
\end{tabular},
\begin{tabular}{|c|}
  \hline
  $\hspace{0.08cm} s\hspace{0.08cm}$ \\
  \hline
  $\hspace{0.08cm} s\hspace{0.08cm}$ \\
  \hline
  $\hspace{0.08cm} s\hspace{0.08cm}$ \\
  \hline
  $\hspace{0.08cm} s\hspace{0.08cm}$ \\
  \hline
  $\hspace{0.08cm} s\hspace{0.08cm}$ \\
  \hline
\end{tabular}
 \right)_{CS}$. \\

\section{$q^7 s^5(I=\frac{1}{2},S=2)$}

1. $\begin{tabular}{|c|c|c|c|}
  \hline
  \quad \quad & \quad \quad & \quad \quad & $\hspace{0.08cm} s\hspace{0.08cm}$ \\
  \hline
  \quad \quad & \quad \quad & $\hspace{0.08cm} s\hspace{0.08cm}$ & $\hspace{0.08cm} s\hspace{0.08cm}$ \\
  \hline
  \quad \quad & \quad \quad & $\hspace{0.08cm} s\hspace{0.08cm}$ & $\hspace{0.08cm} s\hspace{0.08cm}$ \\
  \hline
\end{tabular}_C \otimes$
$\begin{tabular}{|c|c|c|c|c|c|c|c|}
  \hline
  \quad \quad & \quad \quad & \quad \quad & \quad \quad & \quad \quad & \quad \quad &  $\hspace{0.08cm} s\hspace{0.08cm}$ &  $\hspace{0.08cm} s\hspace{0.08cm}$\\
  \hline
  \quad \quad & $\hspace{0.08cm} s\hspace{0.08cm}$ & $\hspace{0.08cm} s\hspace{0.08cm}$ & $\hspace{0.08cm} s\hspace{0.08cm}$  \\
  \cline{1-4}
\end{tabular}_S \rightarrow$
$\left(\begin{tabular}{|c|c|}
  \hline
  \quad \quad & \quad \quad \\
  \hline
  \quad \quad & \quad \quad \\
  \hline
  \quad \quad & \quad \quad \\
  \hline
  \quad \quad \\
  \cline{1-1}
\end{tabular},
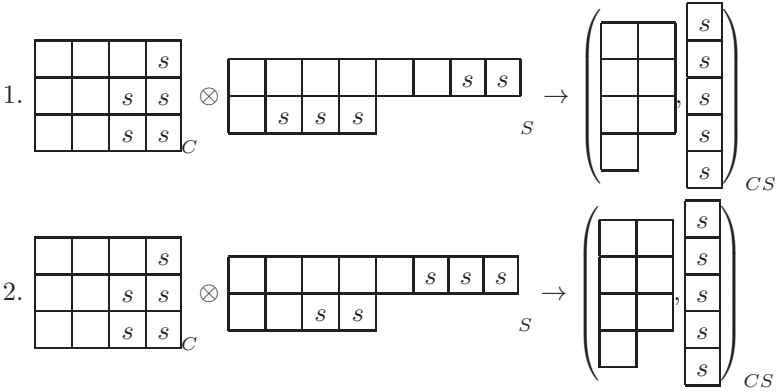
\begin{tabular}{|c|}
  \hline
  $\hspace{0.08cm} s\hspace{0.08cm}$ \\
  \hline
  $\hspace{0.08cm} s\hspace{0.08cm}$ \\
  \hline
  $\hspace{0.08cm} s\hspace{0.08cm}$ \\
  \hline
  $\hspace{0.08cm} s\hspace{0.08cm}$ \\
  \hline
  $\hspace{0.08cm} s\hspace{0.08cm}$ \\
  \hline
\end{tabular}
 \right)_{CS}$. \\

2. $\begin{tabular}{|c|c|c|c|}
  \hline
  \quad \quad & \quad \quad & \quad \quad & $\hspace{0.08cm} s\hspace{0.08cm}$ \\
  \hline
  \quad \quad & \quad \quad & $\hspace{0.08cm} s\hspace{0.08cm}$ & $\hspace{0.08cm} s\hspace{0.08cm}$ \\
  \hline
  \quad \quad & \quad \quad & $\hspace{0.08cm} s\hspace{0.08cm}$ & $\hspace{0.08cm} s\hspace{0.08cm}$ \\
  \hline
\end{tabular}_C \otimes$
$\begin{tabular}{|c|c|c|c|c|c|c|c|}
  \hline
  \quad \quad & \quad \quad & \quad \quad & \quad \quad & \quad \quad & $\hspace{0.08cm} s\hspace{0.08cm}$ &  $\hspace{0.08cm} s\hspace{0.08cm}$ &  $\hspace{0.08cm} s\hspace{0.08cm}$\\
  \hline
  \quad \quad & \quad \quad & $\hspace{0.08cm} s\hspace{0.08cm}$ & $\hspace{0.08cm} s\hspace{0.08cm}$  \\
  \cline{1-4}
\end{tabular}_S \rightarrow$
$\left(\begin{tabular}{|c|c|}
  \hline
  \quad \quad & \quad \quad \\
  \hline
  \quad \quad & \quad \quad \\
  \hline
  \quad \quad & \quad \quad \\
  \hline
  \quad \quad \\
  \cline{1-1}
\end{tabular},
\begin{tabular}{|c|}
  \hline
  $\hspace{0.08cm} s\hspace{0.08cm}$ \\
  \hline
  $\hspace{0.08cm} s\hspace{0.08cm}$ \\
  \hline
  $\hspace{0.08cm} s\hspace{0.08cm}$ \\
  \hline
  $\hspace{0.08cm} s\hspace{0.08cm}$ \\
  \hline
  $\hspace{0.08cm} s\hspace{0.08cm}$ \\
  \hline
\end{tabular}
 \right)_{CS}$. \\

\section{$q^7 s^5(I=\frac{1}{2},S=1)$}

1. $\begin{tabular}{|c|c|c|c|}
  \hline
  \quad \quad & \quad \quad & \quad \quad & $\hspace{0.08cm} s\hspace{0.08cm}$ \\
  \hline
  \quad \quad & \quad \quad & $\hspace{0.08cm} s\hspace{0.08cm}$ & $\hspace{0.08cm} s\hspace{0.08cm}$ \\
  \hline
  \quad \quad & \quad \quad & $\hspace{0.08cm} s\hspace{0.08cm}$ & $\hspace{0.08cm} s\hspace{0.08cm}$ \\
  \hline
\end{tabular}_C \otimes$
$\begin{tabular}{|c|c|c|c|c|c|c|}
  \hline
  \quad \quad & \quad \quad & \quad \quad & \quad \quad & \quad \quad & $\hspace{0.08cm} s\hspace{0.08cm}$ &  $\hspace{0.08cm} s\hspace{0.08cm}$ \\
  \hline
  \quad \quad & \quad \quad & $\hspace{0.08cm} s\hspace{0.08cm}$ & $\hspace{0.08cm} s\hspace{0.08cm}$ &  $\hspace{0.08cm} s\hspace{0.08cm}$  \\
  \cline{1-5}
\end{tabular}_S \rightarrow$
$\left(\begin{tabular}{|c|c|}
  \hline
  \quad \quad & \quad \quad \\
  \hline
  \quad \quad & \quad \quad \\
  \hline
  \quad \quad & \quad \quad \\
  \hline
  \quad \quad \\
  \cline{1-1}
\end{tabular},
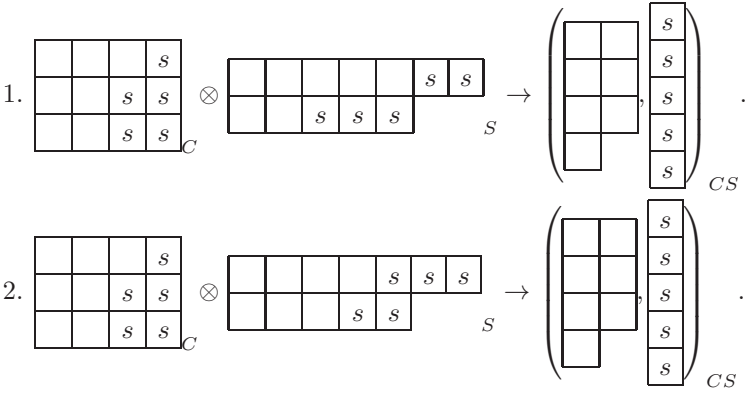
\begin{tabular}{|c|}
  \hline
  $\hspace{0.08cm} s\hspace{0.08cm}$ \\
  \hline
  $\hspace{0.08cm} s\hspace{0.08cm}$ \\
  \hline
  $\hspace{0.08cm} s\hspace{0.08cm}$ \\
  \hline
  $\hspace{0.08cm} s\hspace{0.08cm}$ \\
  \hline
  $\hspace{0.08cm} s\hspace{0.08cm}$ \\
  \hline
\end{tabular}
 \right)_{CS}$. \\

2. $\begin{tabular}{|c|c|c|c|}
  \hline
  \quad \quad & \quad \quad & \quad \quad & $\hspace{0.08cm} s\hspace{0.08cm}$ \\
  \hline
  \quad \quad & \quad \quad & $\hspace{0.08cm} s\hspace{0.08cm}$ & $\hspace{0.08cm} s\hspace{0.08cm}$ \\
  \hline
  \quad \quad & \quad \quad & $\hspace{0.08cm} s\hspace{0.08cm}$ & $\hspace{0.08cm} s\hspace{0.08cm}$ \\
  \hline
\end{tabular}_C \otimes$
$\begin{tabular}{|c|c|c|c|c|c|c|c|}
  \hline
  \quad \quad & \quad \quad & \quad \quad & \quad \quad & $\hspace{0.08cm} s\hspace{0.08cm}$ & $\hspace{0.08cm} s\hspace{0.08cm}$ &  $\hspace{0.08cm} s\hspace{0.08cm}$ \\
  \hline
  \quad \quad & \quad \quad & \quad \quad & $\hspace{0.08cm} s\hspace{0.08cm}$ &  $\hspace{0.08cm} s\hspace{0.08cm}$  \\
  \cline{1-5}
\end{tabular}_S \rightarrow$
$\left(\begin{tabular}{|c|c|}
  \hline
  \quad \quad & \quad \quad \\
  \hline
  \quad \quad & \quad \quad \\
  \hline
  \quad \quad & \quad \quad \\
  \hline
  \quad \quad \\
  \cline{1-1}
\end{tabular},
\begin{tabular}{|c|}
  \hline
  $\hspace{0.08cm} s\hspace{0.08cm}$ \\
  \hline
  $\hspace{0.08cm} s\hspace{0.08cm}$ \\
  \hline
  $\hspace{0.08cm} s\hspace{0.08cm}$ \\
  \hline
  $\hspace{0.08cm} s\hspace{0.08cm}$ \\
  \hline
  $\hspace{0.08cm} s\hspace{0.08cm}$ \\
  \hline
\end{tabular}
 \right)_{CS}$. \\

\end{widetext}

\section*{Acknowledgments}
This work was supported by the Korea National Research Foundation under the grant number 2021R1I1A1A01043019.

\end{document}